\documentclass[acmsmall,screen,usenames,dvipsnames,svgnames,table,final
]{acmart}
\acmJournal{PACMPL}
\acmVolume{1}
\acmNumber{ECOOP}
\acmArticle{1}
\acmYear{2023}
\acmMonth{1}
\acmDOI{} \startPage{1}

\AtBeginDocument{\providecommand\BibTeX{{\normalfont B\kern-0.5em{\scshape i\kern-0.25em b}\kern-0.8em\TeX}}}

\setcopyright{none}

\usepackage{hyperref}
\usepackage{amsmath}
\usepackage{mathtools}
\usepackage{extarrows}
\usepackage{thm-restate}
\usepackage{graphicx}
\usepackage{ifthen}
\usepackage{tikz}
\usetikzlibrary{arrows,automata,calc}
\usetikzlibrary{shapes.symbols,shadows,decorations}
\tikzset{
  gt/.style={
         ->,
         >=stealth',
         shorten >=.1pt,
         auto,
         node distance=5cm,
         scale = 1,
         every state/.style={inner sep = 2pt, minimum size = 0pt, font=\footnotesize},
         transform shape
  },
lt/.style={
	 ->,>=stealth',shorten >=1pt,auto,node distance=5cm,scale = 0.97,transform shape,
  },
  initial/.style={
	 state,initial by arrow, initial text={}
  }
}
\usepackage[capitalise]{cleveref}
\usepackage{xargs}

\usepackage{listings}
\lstdefinelanguage{JavaScript}{
  morekeywords=[1]{break, continue, delete, else, for, function, if, in,
    new, return, this, typeof, var, void, while, with, type, constructor, public},
morekeywords=[2]{false, null, true, boolean, number, undefined,
    Array, Boolean, Date, Math, Number, String, Object},
morekeywords=[3]{eval, parseInt, parseFloat, escape, unescape},
  sensitive,
  morecomment=[s]{/*}{*/},
  morecomment=[l]//,
  morecomment=[s]{/**}{*/}, morestring=[b]',
  morestring=[b]"
}[keywords, comments, strings]

\lstdefinelanguage[ECMAScript2015]{JavaScript}[]{JavaScript}{
  morekeywords=[1]{await, async, case, catch, class, const, default, do,
    enum, export, extends, finally, from, implements, import, instanceof,
    let, static, super, switch, throw, try},
  morestring=[b]` }
\lstalias[]{ES6}[ECMAScript2015]{JavaScript}

\definecolor{mediumgray}{rgb}{0.3, 0.4, 0.4}
\definecolor{mediumblue}{rgb}{0.0, 0.0, 0.8}
\definecolor{forestgreen}{rgb}{0.13, 0.55, 0.13}
\definecolor{darkviolet}{rgb}{0.58, 0.0, 0.83}
\definecolor{royalblue}{rgb}{0.25, 0.41, 0.88}
\definecolor{crimson}{rgb}{0.86, 0.8, 0.24}

\lstdefinestyle{JSES6Base}{
  basicstyle=\ttfamily\footnotesize\lstextra,
  breakatwhitespace=false,
  breaklines=false,
  columns=fullflexible,
  commentstyle=\color{mediumgray}\upshape,
  emph={},
  emphstyle=\color{crimson},
  extendedchars=true,  fontadjust=true,
  identifierstyle=\color{black},
  keepspaces=true,
  keywordstyle=\color{mediumblue},
  keywordstyle={[2]\color{darkviolet}},
  keywordstyle={[3]\color{royalblue}},
  numbers=left,
  numbersep=5pt,
  numberstyle=\tiny\color{black},
  showlines=true,
  showspaces=false,
  showstringspaces=false,
  showtabs=false,
  stringstyle=\color{forestgreen},
  tabsize=2,
  title=\lstname,
  upquote=true  }
\let\lstextra=\relax

\lstdefinestyle{JavaScript}{
  language=JavaScript,
  style=JSES6Base
}
\lstdefinestyle{ES6}{
  language=ES6,
  style=JSES6Base
}

\usepackage{mfirstuc}

\usepackage{xspace}

\usepackage{todonotes}

\newif\ifemi

\def\finex{{\unskip\nobreak\hfil
\penalty50\hskip1em\null\nobreak\hfil$\diamond$
\parfillskip=0pt\finalhyphendemerits=0\endgraf}}

\newcommand{\eMcomm}[2][check]{\ifthenelse{\equal{#1}{new}}{{\color{red}#2}}{\ifthenelse{\equal{#1}{changed}}{{\color{teal}{#2}}}{\todo[color=orange!20]{\tiny eM: \color{NavyBlue}#1}{\color{OliveGreen}{#2}}}}}
\newcommand{\hercomm}[2][check]{
  \ifthenelse{\equal{#1}{new}}{{\color{magenta}#2}}{
    \todo[color=orange!20]{\tiny HM: \color{NavyBlue}#1}
{\color{OliveGreen}{#2}}
}
}

\newcommand{\tnxbehapi}[1][partly]{Research {#1} supported by the EU H2020 RISE programme under the
  Marie Sk{\l}odowska-Curie grant agreement No 778233}

\newcommand{\tnxitmatters}[1][Work partially funded]{#1 by MIUR project PRIN 2017FTXR7S \emph{IT MATTERS}
  (Methods and Tools for Trustworthy Smart Systems)}

\newcommand{\tnxtardis}{\texttt{machine-runner} and \texttt{machine-check} partly funded by the European Union (TaRDIS, 101093006)}

\DeclareGraphicsExtensions{.png,.PNG,.pdf,.PDF,.jpg,.mps,.jpeg,.jbig2,.jb2,.JPG,.JPEG,.JBIG2,.JB2}

\usepackage{bm}
\usepackage{fixme}
\fxusetheme{color}
\FXRegisterAuthor{eM}{aeM}{\color{orange} {\underline{eM}}}

\usepackage[normalem]{ulem} 

\usepackage{xifthen}        \newcommand{\ifempty}[3]{\ifthenelse{\isempty{#1}}{#2}{#3}}

\newcommand{\mkfun}[4][\colorFun]{
  \newcommand{#2}[1][#4]{
    {#1\textit{#3}}
    \ifempty{##1}{}{\,({##1})}
  }
}

\newcommand \hd [1]{{#1[0]}}

\newcommand{\hidden}[1]{}

\newcommand{\cf}[2]{
  \fontsize{#1}{#1}{\selectfont{#2}}
}
\ifemi
\usepackage{showlabels}

\newcommand{\emi}[2]{
  \marginpar{\fcolorbox{red}{shadecolor}{\cf{#1}{{#2}}}}
}
\newcommand{\emic}[2]{\par
  \fcolorbox{red}{shadecolor}{\parbox{\linewidth}{
      \color{gray}
      \begin{description}
      \item[{\color{blue} #2}]{\sf #1}
      \end{description}}}
}
\else
\newcommand{\emi}[2]{}
\newcommand{\emic}[2]{}{}
\fi

\usepackage{mathbbol}

\newcommand{\sst}{\;\big|\;}
 \newcommand{\dom}[1]{\operatorname{dom} {#1}}

\newcommand{\conf}[1]{\ensuremath{\langle {#1} \rangle}}

\newcommand{\bnfdef}{\ ::=\ }

\newcommand{\qqand}[1][and]{\qquad\text{#1}\qquad}
\newcommand{\qand}[1][and]{\quad\text{#1}\quad}

\newcommand{\nat}{\mathbb{N}}
\newcommand{\upd}[3]{{#1}[{#2} \mapsto {#3}]}

{\bfseries}{\rmfamily}
{\bfseries}{\rmfamily}

\newcommand{\quo}[1]{\lq\lq {#1}\rq\rq}
\def\finex{{\unskip\nobreak\hfil
\penalty50\hskip1em\null\nobreak\hfil$\diamond$
\parfillskip=0pt\finalhyphendemerits=0\endgraf}}

\definecolor{shadecolor}{rgb}{1,0.99,0.9}
\definecolor{bg}{rgb}{0.95,0.95,0.95}

\definecolor{myblue}{rgb}{0,0,0.6}
\definecolor{myred}{rgb}{0.4,0,0}
\definecolor{otherred}{rgb}{0.5,0,0.2}

\definecolor{mygreen}{rgb}{0.3,0.4,0.3}

\def\colorSet{\color{otherred}}
\def\colorFish{\color{myred}}
\def\colorVar{\color{myred}}
\def\colorElem{\color{otherred}}
\def\colorCmd{\color{myblue}}
\def\colorLog{\color{otherred}}
\def\colorFun{\color{mygreen}}
\def\colorSymb{\color{mygreen}}

\DeclareMathOperator{\branchsep}{{\&}}
\DeclareMathOperator{\logcat}{{\colorSymb\cdot}}

\DeclareMathOperator{\sublog}{{\colorFun \sqsubseteq}}
\DeclareMathOperator{\mergelog}{{\colorFun \bowtie}}
\DeclareMathSymbol{\mdotsymb}{\mathord}{symbols}{"01}
\DeclareMathOperator{\mdot}{\!{\colorSymb \mdotsymb}\!}
\DeclareMathOperator{\cte}{{\colorSymb /}}

\makeatletter
\newcommand{\dolist}[2]{\def\nextitem{\def\nextitem{#1}}\@for \el:=#2\do{\nextitem\el}}
\makeatother

\newcommand{\wf}[1][$\reactsto$]{{#1}-WF}
\newcommandx{\realise}[2][1=\reactsto,2=\agt,usedefault=@]{$(#1,#2)$-realisation}

\newcommand{\natseg}[1][n]{\underline{#1}}
\newcommand{\eventset}{{\colorSet{\mathcal{E}}}}
\newcommand{\typeset}{{\colorSet{\mathcal{T}}}}

\newcommandx{\aevent}[2][1={},2=e,usedefault=@]{
  {\colorElem\mathit{\MakeLowercase{#2}}}_{#1}
}
\newcommandx{\acevent}[3][1={},2=e, 3=S, usedefault=@]{({#3}_{#1},\aevent[][{#2}]_{#1})}
\newcommandx{\aclog}[2][1={},2=\lambda,usedefault=@]{{\colorElem {#2}}_{#1}}

\newcommandx{\aeventtype}[2][1={},2=t,usedefault=@]{
\mathsf{\colorElem {{#2}}}_{\colorElem #1}
}
\newcommandx{\eventtyping}[3][1={}, 2=e, 3=t, usedefault=@]{
  \vdash \aevent[{#1}][{#2}] \, {\colorSymb :}\, \aeventtype[{#1}][{#3}]
}
\newcommand{\emptylog}{\epsilon}

\newcommandx{\eqg}[3][1=\aclog,2=\aclog',3=\agt]{{#1}\equiv_{#3}{#2}}

\newcommandx{\alogtype}[2][1={},2=l,usedefault=@]{{{\colorLog\mathtt{#2}}_{#1}}}
\newcommandx{\alog}[2][1={},2=l,usedefault=@]{
  {\colorLog\mathit{#2}}_{#1}
}
\newcommandx{\logof}[1][1={},usedefault=@]{\mathsf{\colorFun log}\ifempty{#1}{}{({#1})}}
\newcommandx{\reachable}[1][1={},usedefault=@]{\mathsf{\colorFun Reach}\ifempty{#1}{}{({#1})}}
\newcommand{\source}{{\colorFun source}}
\newcommand{\subterms}{{\colorFun subterms}}

\newcommandx{\efftype}[4][1=\alog,2=\agt,3=\reactsto,4={},usedefault=@]{
\mathit{\colorFun \mathbb{T}}_{#3}\ifempty{#4}{({#1},{#2})}{({#1},{#2},{#4})}}

\newcommandx{\eqlog}[3][3=\aspecdef,usedefault=@]{#1\equiv_{#3}#2}

\newcommandx{\leqlog}[1][1=\alog]{\, <_{#1} \,}
\newcommandx{\geqlog}[1][1=\alog]{\, >_{#1} \,}

\newcommandx{\acmd}[2][1={},2=c,usedefault=@]{{\colorCmd\mathsf{{#2}}}_{#1}}
\newcommand{\inputsymbol}{{\colorSymb ?}}

\newcommandx{\apref}[2][1={},2=e,usedefault=@]{\aeventtype[{#1}][{#2}]\!\inputsymbol}
\newcommandx{\inp}[2][1={}, 2=t, usedefault=@]{
  \aeventtype[{#1}][{#2}]\inputsymbol\,
}
\makeatletter
\newcommandx{\inps}[2][1={}, 2=t, usedefault=@]{
  \def\nextitem{\def\nextitem{#1}}\@for \el:=#2 \do{\aeventtype[{#1}][{\el}]\inputsymbol\, \nextitem}}
\makeatother

\newcommandx{\outp}[2][1=\acmd,2=\aeventtype]{#1/{{#2}\colorSymb{!}}}

\newcommand{\aC}{\colorElem{C}}
\newcommand{\acmdrel}{{\colorElem{\kappa}}}

\newcommandx{\afish}[4][1={},2=M, 3=\aC, 4=\acmdrel, usedefault=@]{\ifempty{#2}{{#3}_{#1} \cdot {#4}_{#1}} {{{\colorFish \mathtt{#2}}_{#1}}}}
\newcommandx{\afishact}[3][1={},2=c,3=,usedefault=@]{
  \acmd[{#1}][{#2}] \cte \ifempty{#3}{\alogtype[{#1}]}{\aeventtype[{#1}][{#3}]}
}
\newcommandx{\obscmd}[3][1=\afish, 2={\acmd[]}, 3 = {\alogtype[]}, usedefault=@]{
  {\big(#1\big)}\!\downarrow_{\afishact[][{#2}][{#3}]}
}

\newcommandx{\agt}[2][1={},2=G,usedefault=@]{\ifempty{#1}{{\colorFish \mathsf{#2}}}{{\colorFish \mathsf{#2}}_{#1}}}

\newcommandx{\fishid}[1][1=f]{{\colorFish {\mathtt{#1}}}}

\newcommandx{\asys}[2][1={},2=S,usedefault=@]{{{\colorElem\mathbb{#2}}_{#1}}}
\newcommandx{\apond}[2][1={},2=M,usedefault=@]{{{\color{myred}\mathcal{#2}}_{#1}}}
\newcommand{\logset}{{\color{myred}\mathcal{L}}}
\newcommand{\roleset}{{\color{myred}\mathcal{R}}}

\newcommandx{\asysrt}[2][1={},2=S,usedefault=@]{{{\color{myred}\mathtt{#2}}_{#1}}}
\newcommandx{\metasys}[2][1={},2=S,usedefault=@]{{{\colorElem\mathfrak{#2}}_{#1}}}

\newcommandx{\fishes}[2][1={},2=L,usedefault=@]{{\colorElem\mathbb{#2}}_{#1}}

\newcommandx{\arsys}[2][1={},2=S,usedefault=@]{{\colorElem{\mathbb {#2}_{#1}}}}

\newcommandx{\areaction}[3][1=i,2=t,3=\afish,usedefault=@]{
  \aeventtype[{#1}][{#2}] {\colorSymb ?}\, {{#3}_{#1}}
}

\makeatletter
\def\mklogtype#1{
  \def\nextitem{\def\nextitem{\logcat\penalty0}}\@for\el:=#1\do{\nextitem\aeventtype[][{\el}]}}

\def\mklog#1{
  \ifempty{#1}{\emptylog}{
    \def\nextitem{\def\nextitem{\logcat\penalty0}}\@for\el:=#1\do{\nextitem{\el}}}
}

\def\mksys#1{
  \def\transform##1[##2]{(\afish[][##1],\mklog{##2})\parop\penalty0}\@for\el:=#1\do{\expandafter\transform\el}}
\makeatother

\newcommandx{\asumnew}[2][1=\acmdrel,2={\areaction[]},usedefault=@]{
  \def\tmp{\dolist{\ \branchsep\ }{#2}}
  \ifempty{#1}{\tmp}{{#1} \mdot \ifempty{#2}{\zero}{\left[ \tmp \right]}}
}
\newcommand{\atrole}{\texttt{\colorSymb \upshape@}}

\newcommandx{\gsumpref}[5][1={},2=c,3=t,4=R,5={}]{\acmd[{#1}][{#2}] \cte \alogtype[{#1}][{#3}] \atrole \arole[{#1}][{#4}]{\gtpref}\ifempty{#5}{\zero}{{#5}_{#1}}}

\newcommand{\gtpref}{\,{\colorSymb .}\penalty0\,}

\newcommandx{\gsumprefix}[4][1=i,2=c,3=l,4=R,usedefault=@]{
\acmd[{#1}][{#2}]{\atrole\arole[{#1}][{#4}]}\conf{{\def\rkl{l}\def\rkx{#3}\ifx\rkl\rkx\alogtype[{#1}][{#3}]\else\mklogtype{#3}\fi}}
}

\newcommandx{\arole}[2][1=i,2=R,usedefault=@]{{\mathtt{#2}_{#1}}}
\newcommandx{\agsum}[6][1=i,2=I,3=\acmd,4=l,5=G,6={\arole[]},usedefault=@]{
    \sum_{#1 \in #2}\gsumprefix[{#1}][{#3}][{#4}][{#6}] \gtpref \agt[{#1}][{#5}]}

\newcommandx{\agprodrun}[5][1=i,2=I,3=\acmd,4=\alogtype,5=\agt,usedefault=@]{
    \pi_{#1 \in #2}{#3_#1/#4_#1;#5}}

\mkfun[\colorFun]{\reactsto}{$\sigma$}{}
\mkfun[\colorFun]{\achi}{$\chi$}{}
\newcommand{\alambda}{{\reactsto}}

\newcommandx{\aspecdef}[3][1=\agt,2=\achi,3=\reactsto,usedefault=@]{#1,#3}
\mkfun[\colorFun]{\cmds}{cmds}{}
\mkfun[\colorFun]{\events}{events}{}

\mkfun[\colorFun]{\norec}{norec}{}
\mkfun[\colorFun]{\auxcr}{cr}{}
\mkfun[\colorFun]{\cmdrdy}{crdy}{}
\mkfun[\colorFun]{\rdy}{guards}{}
\mkfun[\colorFun]{\cmpts}{roles}{}
\mkfun[\colorFun]{\roles}{roles}{}
\mkfun[\colorFun]{\filter}{filter}{}
\mkfun[\colorFun]{\activer}{active}{}
\mkfun[\colorFun]{\passiver}{passive}{}
\mkfun[\colorFun]{\observed}{obs}{}

\newcommandx{\aproj}[3][1={\arole[]},2=\reactsto,3=\achi,usedefault=@]{\downarrow_{#1}^{{#2}}}

\newcommandx{\asem}[1][1={}]{[\![ #1 ]\!]}

\newcommandx{\atripg}[3][1=\agt,2=\alambda,3=\achi]{\langle#1,#2,#3\rangle}
\renewcommand{\atripg}{\agt}

\newcommand{\zero}{{{\colorSymb \mathbf{0}}}}
\newcommandx{\avar}[1][1=X]{{\colorVar \mathtt{#1}}}
\newcommandx{\avartype}[1][1=X]{{\colorVar \mathtt{#1}}}

\newcommandx{\arec}[2][1=\avar,2=\afish,usedefault=@]{
  {{#1} = {#2}}
}

\newcommand{\parop}{\,{\colorSymb \mid }\,}
\newcommand{\stchange}[1][]{{\colorFun{\delta}_{#1}}}
\newcommand{\Stchange}[1][]{{\colorFun{\delta}_{#1}}}

\newcommandx{\red}[1][1={},usedefault=@]{\xlongrightarrow{#1}}
\newcommandx{\weakred}[1][1={},usedefault=@]{\xLongrightarrow{#1}}
\newcommandx{\absred}[1][1={},usedefault=@]{\xmapsto{#1}}
\makeatletter
\def\rightarrowfill@@{\arrowfill@@\relax\relbar\rightarrow}
\def\arrowfill@@#1#2#3#4{$\m@th\thickmuskip0mu\medmuskip\thickmuskip\thinmuskip\thickmuskip
   \relax#4#1
   \xleaders\hbox{$#4#2$}\hfill
   #3$}
 \newcommand{\metared}[2][]{\ext@arrow 0359\rightarrowfill@@{#1}{#2}}
\makeatother

\newcommand{\irule}[2]{\frac{\textstyle\rule[-1.3ex]{0cm}{3ex}#1}{\textstyle\rule[-.5ex]{0cm}{3ex}#2}}

\newcommand{\rulename}[1]{[{\sc#1}]}

\def \mathrule #1#2#3{
    \irule{#1}{#2}\ifempty{#3}{}{\hspace{0em}\mbox{\footnotesize\rulename{#3}}}
}

\newcommand{\smr}{{\sc smr}\xspace}

\begin{document}

\title{Behavioural Types for Local-First Software}
\thanks{\tnxbehapi.
  \tnxitmatters.
  \tnxtardis.

  The authors also thank the anonymous reviewers for their useful and insightful comments
  and Daniela Marottoli for her help in the initial phase of this project.
}

\author{Roland Kuhn}
\email{roland@actyx.io}
\orcid{0000-0003-1582-6238}
\affiliation{
  \institution{Actyx AG}
  \country{Germay}
}

\author{Hern\'an Melgratti}
\orcid{0000-0003-0760-0618y}
\email{hmelgra@dc.uba.ar}
\affiliation{\institution{Universidad de Buenos Aires \& Conicet}
  \country{Argentina}
}

\author{Emilio Tuosto}
\orcid{0000-0002-7032-3281}
\affiliation{\institution{Gran Sasso Science Institute}
  \country{Italy}
}

\begin{abstract}
  Peer-to-peer systems are the most resilient form of distributed
  computing, but the design of robust protocols for their coordination
  is difficult.
This makes it hard to specify and reason about global behaviour of
  such systems.
This paper presents \emph{swarm protocols} to specify such systems
  from a \emph{global} viewpoint.
Swarm protocols are projected to \emph{machines}, that is
  \emph{local} specifications of peers.
  
  We take inspiration from behavioural types with a key difference:
  peers communicate through an event notification mechanism rather
  than through point-to-point message passing.
Our goal is to adhere to the principles of
  \emph{local-first software} where network devices collaborate on a
  common task while retaining full autonomy:
  every participating device can locally make progress at all times,
  not encumbered by unavailability of other devices or network connections.
This coordination-free approach leads to inconsistencies that may
  emerge during computations.
Our main result shows that under suitable well-formedness conditions
  for swarm protocols consistency is eventually recovered and the
  locally observable behaviour of conforming machines will eventually
  match the global specification.
  
  The model we propose elaborates on an existing industrial platform
  and provides the basis for tool support (sketched here and fully
  described in a companion artifact paper), wherefore we consider this
  work to be a viable step towards reasoning about local-first and
  peer-to-peer software systems.
\end{abstract}

\ccsdesc[500]{Software and its engineering}
\ccsdesc[500]{Software and its engineering~Software notations and tools}

\keywords{Distributed coordination, local-first software, behavioural types, publish--subscribe, asynchronous communication}

\maketitle
\section{Introduction}\label{sec:in}

Fully decentralised systems like peer-to-peer networks are notoriously hard to
design and analyse.
A main challenge is to coordinate components so that the composed system
exhibits the expected behaviour.
As all decisions in such a system are made locally based on the available
partial knowledge, the main problem is to specify which information is
transferred to whom and when and how to interpret it, i.e. protocol design.
We illustrate this using the following scenario.

\begin{example}[Our running example]\label{ex:scenario}\label{ex:local-bidding}
  A taxi fleet organises rides within a town using a peer-to-peer network.
The main goal is to provide any passenger who requests a ride with some
  offers from taxis willing to perform the desired transportation; the
  passenger may pick one offer based on price and estimated time of
  arrival---followed by tracking the pickup, ride, and arrival---or cancel the
  ride.
At the end of the trip the accounting office provides a receipt for the
  journey or cancellation.

  The structure of the expected interaction between the different
  peers (classified by \emph{roles}) can be informally illustrated
  with the following diagram involving a passenger role $\arole[][P]$,
  a taxi role $\arole[][T]$, and an accounting office role as
  $\arole[][O]$.
  \begin{align*}
    \begin{tikzpicture}[gt,font=\footnotesize]
      \def\l#1#2{$\acmd[][#1]\atrole\arole[][#2]$}
      \node[initial] (1) {$1$};
      \foreach \n/\p in {2/1,3/2,4/3,5/4,6/5,7/6,8/7}{ \node[state,right=13mm of \p] (\n) {$\n$}; }
      \path (1) edge node{\l{Request}{P}} (2);
      \path (2) edge node{\l{Offer}{T}} (3);
      \path (3) edge[loop above] node{\l{Offer}{T}} ();
      \path (3) edge node{\l{Select}{P}} (4);
      \path (4) edge node{\l{Arrive}{T}} (5);
      \path (5) edge node{\l{Start}{P}} (6);
      \path (6) edge[loop above] node{\l{Record}{T}} ();
      \path (6) edge node{\l{Finish}{P}} (7);
      \path (4) edge[bend right=22] node[sloped]{\l{Cancel}{P}} (7);
      \path (7) edge node{\l{Receipt}{O}} (8);
    \end{tikzpicture}
  \end{align*}
When a passenger needs a ride (state $1$ in the diagram above), they open an
  auction by executing a $\acmd[][Request]$.
Then, any taxi with capacity can proceed with an  $\acmd[][Offer]$. In
  this description, we do not make any assumption about the number of
  instances playing each role; in fact, we expect to have many taxis playing role $\arole[][T]$ and hence many offers.
The passenger ends the auction by using  $\acmd[][Select]$  to
  pick a winner after at least one offer has been received.

  The second phase starts with a race in state 4: either the taxi
  invokes $\acmd[][Arrive]$ first and the ride begins, or the
  passenger loses patience first and uses the $\acmd[][Cancel]$
  command to back out.
The office will create a receipt in either case, but we must
  settle the dispute whether a ride happened.
\finex \end{example}

The classic solution to such a problem would use a central database to present
an up-to-date view of the respective data each participant is allowed to see.
This solution avoids conflicts, maintains invariants, and steers the whole
process by virtue of there being only one source of truth---one of the
$\acmd[][Arrive]$ or $\acmd[][Cancel]$ commands would happen first and the
other would be rejected. It is well-known (cf.\ the CAP~\cite{CAP} and
FLP~\cite{FLP} results) that such a solution suffers from unavailability if
the system model includes network partitions, since either the database is
localised and may be unreachable, or it is distributed and may need to reject
requests to maintain consistency.
With the advent of \emph{conflict-free replicated data types
  (CRDTs)}~\cite{shapiro2011conflict} a different solution came into view:
instead of avoiding conflicts through coordination, CRDTs provide a data model
of such a structure that is conflict-free by construction. CRDTs facilitate that
by demanding a join semi-lattice for the data structure, i.e. that there is a
merge function that given two differently evolved states will compute a new
state that represents the sum of all operations that were done to either
input---as far as this is possible. Applied to \cref{ex:local-bidding} this
would typically be achieved by systematically preferring either side of the
choice at state 4, e.g. \quo{cancellation always wins} (cf. the \emph{add-wins
  set} CRDT). This illustrates that CRDTs are not well-suited for capturing
and fairly resolving conflicts such as the one in our example.
Nevertheless, the increasing research focus on coordination-free systems
inspired the formulation of \emph{local-first software}~\cite{localfirst} in
which all participants in a distributed system maintain full autonomy and
control over their data. Besides the focus on agency and ownership,
local-first principles can also be used to build software that is fully
available and maximally resilient~\cite{lfc}, where each participant can
independently make decisions that are globally valid.

In this paper we propose a new way of approaching local-first software based
on embracing conflict, recognising it, and finally reconciling it to reach
eventual consensus~\cite{8919675}.
Our model of computation builds upon the middleware developed at
Actyx~\cite{actyx}, which provides a reliable durable pub–sub mechanism for
event logs as well as a coordination-free total order of all events. A
distributed system in our model is realised by a set of participants, dubbed
\emph{machines}, that can exhibit discordant behaviour and interact by
broadcasting and reacting to \emph{events}.
Events are generated locally in response to the execution of \emph{commands},
added to the \emph{local log} and then propagated to other participants, to be
merged into the log local at the recipient.
We assume that events propagate asynchronously and that there is no
traditional mechanism for coordination (like consensus or central nodes):
participating machines liaise with each other purely on the basis of the events
spreading in the system.
We refer to such systems as \emph{swarms}.

Each machine in a swarm implementing the scenario in \cref{ex:scenario}  plays a role, i.e. it subscribes to a defined subset of event types and applies an assigned logic to interpret those.
Depending on its local state (which it computes from its local log), any machine may decide to execute a command.
For instance, a machine playing the role $\arole[][P]$ may execute
$\acmd[][Request]$, which generates new events containing details of the
request. Such events are appended to the local log of the machine and then
propagated asynchronously to other machines that have subscribed to such
events, e.g.\,the machines corresponding to taxis. After receiving such
events, a taxi updates its local state and decides whether to place an offer
for the ride. In such case, it executes $\acmd[][Offer]$, which
generates the events describing the offer, appends them to local log and
propagates them to the rest of the system. The interaction described
in \cref{ex:scenario} may proceed to completion in this way.

It should be noted that our computational model does not preclude the execution of
conflicting commands. In fact, a passenger may cancel the ride while the taxi
may concurrently confirm its arrival (state 4 in \cref{ex:scenario}). In this
case, both machines will generate their respective events and
propagate them through the system. When receiving such events, each
machine will be in charge of detecting and properly resolve the conflict.
This is achieved by using the total order between events---interpreted as
manifestation of (logical) time---to settle conflicts: the earliest event
emitted after a choice decides which branch is taken (the events corresponding
to the losing branch are just ignored). Note that we do not assume knowledge
of when the event log is complete, i.e.\,it is not possible for a machine to
detect whether an event is the earliest possible event after a choice. 
As soon as events up to and including a given choice are fully replicated across the swarm, all machines will agree upon which branch is taken; note that we assume that events are eventually available to all machines.
However, while such events have not yet reached all involved machines, computed local states may temporarily diverge.

The fact that events are processed only by subscribers makes the resolution
of choices subtle. Assume that the accounting office incorrectly subscribes to
the cancellation event but not to the arrival one. In the presence of a
conflicting choice, it may incorrectly conclude that the ride was cancelled
even when all other roles understand that the ride took place.
We rely on a typing discipline for ruling out such inconsistencies in the
resolution of choices.
We follow a top-down approach featuring \emph{swarm protocols}, namely
abstractions similar to the diagram in~\cref{ex:local-bidding} that---akin to global
types~\cite{honda2008multiparty,honda2016multiparty}---formalise a description
of the expected protocol from a \emph{global} viewpoint.
A projection operation can automatically generate local specifications of each
role formally defined as \emph{machines} (cf. \cref{sec:machine-interaction}).
Our typing discipline establishes sufficient
conditions---\emph{well-formedness} of swarm protocols---to guarantee that
well-typed systems will resolve conflicting choices consistently once
information has sufficiently spread to participants.
This and the fact that swarm protocols fully abstract away from the number of
instances enacting a role are distinguished features of our approach.

\smallskip\noindent\textbf{Summary of the main contributions and structure of the paper\ }
We develop a behavioural typing discipline for local-first software tailored to a formal operational model that we distill from a real middleware.
Our theory is also implemented in a toolkit that type checks swarms and enables their simulation. More specifically:
\begin{enumerate}
\item We introduce an operational model for distributed computation based on replication of event logs to drive local state machines (\cref{sec:fishes}). The proposed model does not assume stability but combines speculative computation with a \quo{rewind} mechanism \`a la \emph{time warp machine}~\cite{jefferson1985virtual}: a conflict is resolved by backtracking and re-execution along the right path.
\item We define a novel behavioural type approach (\cref{sec:gt}) in which swarm protocols are specified in terms of the information injected into a heterogeneous swarm through the actions performed by participants of specific roles. Swarm protocols enjoy a lightweight syntax and simple operational semantics; they are deadlock-free and communication-safe (\cref{sec:deadlockfree}); yet they are expressive enough for modelling complex protocols.
\item We define well-formedness conditions for swarm protocols (\cref{def:wf}) and a projection operation to derive local machine specifications (\cref{sec:proj}). These ensure eventual consistency between local observations and globally specified behaviour (\cref{th:main,cor:faithful}), which is non-trivial due to the absence of any infrastructure coordination, non-homogeneous event subscriptions across roles, and the ability to implement a role with an arbitrary positive number of replicas.
\item We apply our approach to the TypeScript language and Actyx middleware in the form of a runtime library and a tool for checking protocol well-formedness and conformance, as well as a stochastic simulation tool exploring possible executions permitted under our model.
\end{enumerate}

The semantics of machines and of log shipping are given in \cref{sec:fishes}, followed by an overview of the accompanying tooling in \cref{sec:tooling}.
We introduce swarm protocols in \cref{sec:gt} followed by their local projection onto machines in \cref{sec:proj}.
We then present the well-formedness conditions in \cref{sec:wf} that we use in \cref{sec:correct} to define our notion of eventual correctness.
\cref{sec:rw} discusses related work while \cref{sec:conc} yields final remarks.
Supplementary material and proofs are relegated to a separate appendix.

\section{Asymmetric Replicated State Machines}\label{sec:fishes}
Our model hinges on three ingredients: machines, event emission and consumption, and log-shipping.
The behaviour of a machine is captured by a finite-state automaton as described in \cref{sec:machine-single}. In \cref{sec:consumption} we show how machines may offer \emph{commands} that, upon execution, emit events as well as how machines consume (typed) events stored in their \emph{local log}. Such events are immediately stored in the local log of the emitting machine and later asynchronously shipped to the other machines as described in \cref{sec:machine-interaction}.

\subsection{From TypeScript to automata}\label{sec:machine-single}

\def\lstextra{\begin{tikzpicture}[gt]
  \useasboundingbox (0,0)rectangle(0,0);
  \node[initial,initial where=right] (1) at (13.3,-2.8) {InitialP};
  \node[state,below=53mm of 1] (2) {AuctionP};
  \node[state,below=30mm of 2] (3) {RideP};
  \path (1)--(2)
    node[state,pos=0.3,anchor=base] (s1) {1}
    node[state,pos=0.6,anchor=base] (s2) {2};
  \node[state,left=2cm of 2,yshift=8mm] (s3) {3};
  \path (2)--(3) node[state,pos=0.7,anchor=base] (s4) {4};
  \path (1) edge[loop left] node[below left]{$\afishact[][Request][Requested]$} ();
  \path (1) edge node[left]{$\inp[][Requested]$} (s1)
        (s1) edge node[left]{$\inp[][Bid]$} (s2)
        (s2) edge node[left,pos=0.3]{$\inp[][BidderID]$} (2);
  \path (2) edge[out=220,in=250,loop] node[left,yshift=2mm]{$\afishact[][Select][Selected\logcat PassengerID]$} ();
  \path (2) edge[bend right,looseness=0.3] node[below,rotate=-20]{$\inp[][Bid]$} (s3)
        (s3) edge[bend right,looseness=0.3] node[below,rotate=-20]{$\inp[][BidderID]$} (2);
		  \path (2) edge node[left,pos=0.7]{$\inp[][Selected]$} (s4)
		  (s4) edge node[left]{$\inp[][PassengerID]$} (3);
\end{tikzpicture}
\gdef\lsextra{\relax}
}
\begin{lstlisting}[style=ES6,float=t,label=lst:ts-passenger,captionpos=b,
    caption={Definition of state machines in TypeScript}]
// analogous for other events; "type" property matches type name (checked by tool)
type Requested = { type: 'Requested'; pickup: string; dest: string }
type Events = Requested | Bid | BidderID | Selected | ...

/** Initial state for role P */
@proto('taxiRide') // decorator injects inferred protocol into runtime
export class InitialP extends State<Events> {
  constructor(public id: string) { super() }
  execRequest(pickup: string, dest: string) {
    return this.events({ type: 'Requested', pickup, dest })
  }
  onRequested(ev: Requested, ev1: Bid, ev2: BidderID) {
    return new AuctionP(this.id, ev.pickup, ev.dest, [{
      price: ev1.price, time: ev1.time, bidderID: ev2.id,
    }])
  }
}
@proto('taxiRide')
export class AuctionP extends State<Events> {
  constructor(public id: string, public pickup: string, public dest: string,
    public bids: BidData[]) { super() }
  onBid(ev1: Bid, ev2: BidderID) {
    const [ price, time ] = ev1
    this.bids.push({ price, time, bidderID: ev2.id })
    return this
  }
  execSelect(taxiId: string) {
    return this.events({ type: 'Selected', taxiID },
                       { type: 'PassengerID', id: this.id })
  }
  onSelected(ev: Selected, id: PassengerID) {
    return new RideP(this.id, ev.taxiID)
  }
}
@proto('taxiRide')
export class RideP extends State<Events> { ... }
\end{lstlisting}

\noindent
We formalise (and elaborate on) the computational model realised in the middleware of Actyx by offering a new library for writing endpoint code.
Like the Actyx SDK, we use the TypeScript language.
Our API focuses on a concise but well-structured expression of finite-state machines for interpreting the current state of distributed computation.
We illustrate this by considering the implementation of the request and auction part of our running taxi example from the passenger's point of view, with the TypeScript code given in Listing~\ref{lst:ts-passenger}.

For the purposes of this section it suffices to know that each state of a machine is represented by a TypeScript class, with methods for command invocation whose name is prefixed with \quo{\texttt{exec}}, and event handler methods to compute the next state (names prefixed with \quo{\texttt{on}}).
The types of the event handler method arguments are significant, as are the return types of command methods.
We describe all other details, including runtime evaluation, in \cref{sec:tooling}.

Listing~\ref{lst:ts-passenger} also depicts the finite state automaton corresponding to the snippet, as inferred by the \texttt{machine-check} build tool (cf. \cref{sec:machine-check}):
\begin{itemize}
\item states \texttt{InitialP}, \texttt{AuctionP}, and
  \texttt{RideP} of the automaton respectively correspond to the
  classes in the snippet with the same name;
\item states 1 and 2 of the automaton correspond to the implicit
  states interspersed between the events specified as arguments to the event handler method
  \texttt{onRequested} (likewise for state 3 from \texttt{onBid} and state 4 from \texttt{onSelected});
\item command methods correspond to self-loops in the automaton, labelled with the command name and the resulting event log type in the form $\afishact$;
\item event handlers correspond to transitions or sequences thereof, where each transition is labelled with an event type of the form $\inp$.
\end{itemize}
The correspondence sketched above is the basis for our formalisation of swarms and it is at the heart of the library \texttt{machine-runner} introduced in \cref{sec:tooling}.

Note that in the automaton we abstract away from payloads, considering only the types of events.
We also ignore internal computations not involving event emission/consumption (e.g.\,the computation of the constructor arguments for state \texttt{AuctionP} is immaterial).

\subsection{Commands execution and events consumption}\label{sec:consumption}

The automata representation of machines discussed in \cref{sec:machine-single} allows us to limit technicalities in defining the behaviour of machines.
(For the reviewer's convenience, the formal definitions of the constructions introduced in this section are found in \cref{sec:formal-machine}.)

Hereafter, we fix a set of event types, ranged over by $\aeventtype[]$, and a set of events, ranged over by $\aevent$. We write $\eventtyping$ when event $\aevent$ has type $\aeventtype$ and assume that this relation is decidable. Each event can be identified and associated with the machine that emitted it: let $\source(\aevent)$ be the identity of the machine generating $\aevent$.
An event log is a sequence of events, written $\mklog{\aevent[1],\aevent[2], \ldots}$.

For the purpose of enabling commands a machine $\afish$ with log $\alog$ is implicitly in a state denoted $\stchange(\afish,\alog)$.
The determinism of $\afish$ ensures that there is a unique such state.
We compute $\stchange(\afish, \alog)$ through a \emph{transition function} by simply adaptating the standard transition function of finite-state automata.
Starting with the initial state of $\afish$ we inductively remove the oldest event, say $\aevent$, and check it against the outgoing transitions of the current state, say $q$: if $q$ has a transition with label $\inp[]$ and $\aevent$ has type $\aeventtype$ we continue with the target state of that transition on the rest of the log, otherwise the event $\aevent$ is dropped and the rest of the log processed from state $q$.

A machine can be thought of as the proxy of an agent (be that an algorithm or a human) that processes the information in the local log, comes to conclusions, and makes decisions which may lead to the invocation of an enabled command. For instance, once in the \texttt{AuctionP} state the machine in \cref{sec:machine-single} enables the passenger to execute the command $\acmd[][Select]$ triggering the emission of an event log like $\mklog{\aevent[][Selected], \aevent[][PassengerID]}$.
This sequence of events is added to the local event log making the machine move to state 2 first by consuming the event $\aevent[][Selected]$ and then to state \texttt{RideP} by consuming the event $\aevent[][PassengerID]$.
This is why the inferred machine state diagram records commands as self-loops while only event consumption may induce state changes.
In the following we only consider deterministic machines, that is those where from each state there are no two different outgoing transitions labelled with the same event type.

It is important to note that the next command may be invoked  only if it is enabled in the state reached after fully processing the local log.
Also, emitted events are \emph{appended} to the local log of the machine they originate at. This ensures causality preservation since the new events are ordered \emph{after} all events previously known by this machine.

Notice that the passenger's machine cannot reach the state \texttt{AuctionP} without consuming an event of type $\aeventtype[][BidderID]$ when it is in state 2. That is, at least one of the machines for the taxis has to put in a bid for the ride. As these machines have different behaviour we call them \emph{asymmetric}.

\subsection{Swarms and log-shipping}\label{sec:machine-interaction}

The last piece to our computational model is how event logs are disseminated among the machines of a swarm.
Once we have established a mechanism for shipping events from one machine to another, the rules of the previous section describe how this affects the recipient: the local log contains more events, leading to a new current state being computed, which in turn may change the set of available commands.

Key to our goal is that different replicas of a same machine reach the same state when consuming the same events.
With our state transition function as defined above this can in general only be achieved by having the events ordered in the same way in the local logs.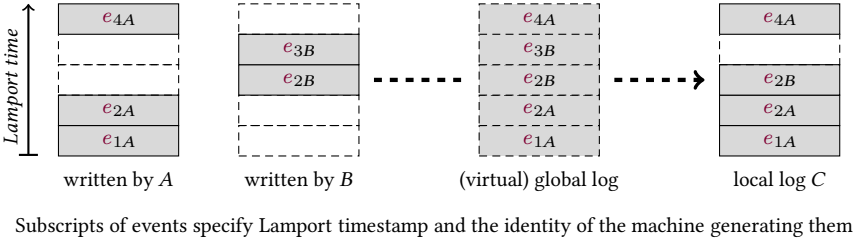
\begin{figure}[t]
  \begin{tikzpicture}[scale=0.4]
    \definecolor{light}{gray}{0.85}
    \tikzstyle{full} = [solid,fill=light,font=\footnotesize]
    \tikzstyle{empty} = [densely dashed]
    \tikzstyle{global} = [densely dashed,fill=light,font=\footnotesize]
	 \tikzstyle{caption} = [align=center,anchor=base, font=\footnotesize]
    \foreach \s/\l/\p in {full/1A/0,full/2A/1,empty/e/2,empty/e/3,full/4A/4}
      \draw[style=\s] (0,\p) +(2,0.5) node{\ifthenelse{\equal\l{e}}{}{$\aevent[\l]$}} (0,\p) rectangle +(4,1);
    \foreach \s/\l/\p in {empty/e/0,empty/e/1,full/2B/2,full/3B/3,empty/e/4}
      \draw[style=\s] (6,\p) +(2,0.5) node{\ifthenelse{\equal\l{e}}{}{$\aevent[\l]$}} (6,\p) rectangle +(4,1);
    \foreach \l/\p in {1A/0,2A/1,2B/2,3B/3,4A/4}
      \draw[style=global] (14,\p) +(2,0.5) node{\ifthenelse{\equal\l{e}}{}{$\aevent[\l]$}} (14,\p) rectangle +(4,1);
    \foreach \s/\l/\p in {full/1A/0,full/2A/1,full/2B/2,empty/e/3,full/4A/4}
      \draw[style=\s] (22,\p) +(2,0.5) node{\ifthenelse{\equal\l{e}}{}{$\aevent[\l]$}} (22,\p) rectangle +(4,1);
    \draw[dashed,->,ultra thick] (10.5,2.5)--(13.5,2.5) (18.5,2.5)--(21.5,2.5);
    \draw[thick,|->] (-1,0)--(-1,5) node[pos=0.9,label={[rotate=90]180:\footnotesize\itshape Lamport time}]{\strut\hskip1ex\strut};
	 \def\vpos{-1}
    \path (2,\vpos) node[caption]{written by $A$}
      (8,\vpos) node[caption]{written by $B$}
      (16,\vpos) node[caption]{(virtual) global log}
      (24,\vpos) node[caption]{local log $C$};
		\path (12.5,-2.5) node[caption]{Subscripts of events specify Lamport timestamp and the identity of the machine generating them};
  \end{tikzpicture}
  \caption{Events sliced by their source: each event starts out at the machine where it is emitted. Logs are disseminated such that the recipient (like machine $C$) holds a prefix of each of the source slices, which is a partial view of the global log. The recipient’s local log is ordered like the global log. Eventually every event arrives at all machines, filling the transient gaps that may have existed.}
  \label{fig:event-log-merging}
\end{figure}
We address this by assuming a total order between events, without coordination between machines (implementability is demonstrated in \cref{sec:actyx}).
One crucial property of this total order is that it must preserve causality: a fresh event goes after its predecessors in the local log since it results from an effect based on the whole local log.
Note that the ordering relative to events that are not yet locally known is arbitrary but well-defined, for which we introduce the conceptual notion of a global log.
Fig~\ref{fig:event-log-merging} illustrates how logs are disseminated: each event starts out in the local log of the machine that emitted it, and at the same time it has its place in the virtual global log, as per the total order.
Log-shipping is modelled by a machine enlarging its replicated subset of the global log; in practice, events are sent from one local log to another, but the precise algorithm for picking source and destination is not important to our theory.

Due to the uncoordinated total order it may happen that an incoming event is sorted into the local log somewhere in the middle, thereby changing the interpretation of all following events in the computation of the current state.
As an example consider that the passenger selects a taxi concurrently to another taxi placing a new bid. If the selection is ordered before the bid, later inspection of the log may reveal that the passenger selected suboptimally---but the selection remains in effect nevertheless, ignoring the bid. In a system based on a central database this corresponds to the case where the \quo{bid} transaction is rejected since it is ordered after the \quo{select} transaction.

If on the other hand the $\aevent[][Selected]$ event were ordered between the $\aevent[][Bid]$ and $\aevent[][BidderID]$ events, it would be ignored once the logs are replicated.
In this case there are two reasonable paths forward: honouring the passenger's wish would require a \emph{compensating action} of executing the selection again, or the new bid should be factored into the decision, possibly leading to a different outcome.

Owing to the way Actyx replicates events, we define the following relation:

\begin{definition}[Sublog relation]\label{def:sublog}
  A log $\alog = \aevent[1] \cdots \aevent[n]$ induces a total order $<_{\alog}$ on its elements as follows:
  $\aevent[i] < _{\alog}\aevent[j] \iff i< j$.
  The \emph{sublog} relation on logs $\sublog$ demands an order-preserving and downward-complete morphism from $\alog[1]$ into $\alog[2]$.
  Formally, $\alog[1] \sublog \alog[2]$ if
  \begin{enumerate}
  \item all events of $\alog[1]$ appear in $\alog[2]$ ($\alog[1] \subseteq \alog[2]$)
    in the same order ($\leqlog[{\alog[1]}] \subseteq \leqlog[{\alog[2]}]$); and
  \item the per-source partitions of \(\alog[1]\) are prefixes of the corresponding partitions of \(\alog[2]\), i.e.
    for all \(\aevent[1]\in\alog[1],\aevent[2]\in\alog[2]\) from a given \(\source\),
    \(\aevent[2] \leqlog[{\alog[2]}] \aevent[1]\) implies \(\aevent[2]\in\alog[1]\).
  \end{enumerate}
\end{definition}

\begin{example}[Sublogs]\label{ex:sublogs}
  Let
  $\alog = \mklog{\aevent[0][a], \aevent[1][b], \aevent[1][d],
	 \aevent[2][c], \aevent[1][e]}$ be a log where events with different subscripts
  are emitted by different machine (i.e.
  $\source(\aevent[0][a]) \neq \source(\aevent[1][b])$). We have
  \(\aevent[1][b]\mdot\aevent[2][c] \sublog \alog\) but not
  \(\aevent[1][b]\mdot\aevent[0][a]\) (due to inverted order) nor
  \(\aevent[1][d]\mdot\aevent[2][c]\) (because the presence of
  \(\aevent[1][d]\) requires that \(\aevent[1][b]\) must also be
  present).
\finex
\end{example}

With this notion in place, our formalisation of log-shipping proceeds in several steps.
First, we model the merging of two logs such that the result contains the union of both sets of events as well as of the $<_{\alog}$ relations;
note that conflicts are impossible because a given event can only appear in both logs due to an earlier merge operation.
Second, we express that global order is arbitrary by noting that the first step is satisfied by a set of possible results given by a shuffling operator\footnote{Our $\mergelog$ operator is a particular case of the synchronous shuffling operator in the words of~\cite{DBLP:journals/jcss/SulzmannT19}, which is known also as \emph{synchronous composition}~\cite{garg1992concurrent} and \emph{produit de mixage}~\cite{de1984languages}.
} $\mergelog$:
\begin{align}\label{eq:merge}
  \alog[1] \mergelog \alog[2] =
  \{\alog \sst \alog \subseteq \alog[1] \cup \alog[2] \qand \alog[1] \sublog \alog \qand \alog[2] \sublog \alog\}
\end{align}
Third, we fix a non-deterministic selection of one possible ordering by demanding that each fresh event is simultaneously added to the virtual global log, which is the source for replication.

\subsection{Formalisation}\label{sec:formalisation}
A \emph{swarm} (of size $n$) is a pair $(\asysrt, \alog)$ where
$\asysrt$ maps indices $1 \leq i \leq n$ to machines and their local
log, i.e. $\asysrt(i) = (\afish[i],\alog[i])$ and $\alog$ is the
global log.
It is convenient to let $(\afish[1],\alog[1]) \parop \ldots \parop (\afish[n],\alog[n]) \parop \alog$ denote the swarm $(\asysrt,\alog)$
such that $\asysrt(i) = (\afish[i],\alog[i])$ for all $1 \leq i \leq n$.
A swarm $(\afish[1],\alog[1]) \parop \ldots \parop (\afish[n],\alog[n]) \parop \alog$ is \emph{coherent} when
$\bigcup_{1 \leq i \leq n}\alog[i] = \alog$ and $\alog[i] \sublog \alog$ for $1 \leq i \leq n$; in other words, the local log $\alog[i]$ of machine $\afish[i]$ is made of
events actually emitted in $\asysrt$ in the order they have been produced.

The type of a log is the sequence of the types of its events. Hereafter, we let $\alogtype$ range on \emph{log types}, that is finite sequences of event types. Accordingly, we write $\eventtyping[][\alog][\alogtype]$ when $\alog$ has type $\alogtype$. Note that the decidability of this typing relation immediately follows from our assumption that the typing of events is decidable.

The inference rules \textsc{Local} and \textsc{Prop} below formalise the behaviour of swarms.
One noteworthy difference to other behavioural typing approaches is that we do not evolve our machine types as the computation progresses, instead we evolve the local and global logs.
Recall that a machine $\afish$ with a log $\alog$ is implicitly in state $\stchange(\afish,\alog)$. We can hence define the relation $(\afish,\alog) \red[{\afishact[][@][\alogtype]}] (\afish,\alog\logcat\alog')$ holding when the state
$\stchange(\afish,\alog)$ enables the command $\afishact[][@][\alogtype]$ and $\alog'$ has type $\alogtype$.
\begin{align*}
  &&&
	 \mathrule{
	 \asysrt(i) = (\afish,\alog[i])
	 \qquad
	 (\afish,\alog[i]) \red[{\afishact[][@][\alogtype]}] (\afish,\alog[i]')
	 \qquad
	 \source({\alog[i]'\setminus\alog[i]})= \{i\}
	 \qquad
	 \alog' \in \alog \mergelog \alog[i]'
	 }{
	 (\asysrt, \alog) \red[{\afishact[][@][\alogtype]}] (\upd \asysrt i {(\afish, \alog[i]')},  \alog')
	 }{Local}
  \\[2em]
  &&&
  \mathrule{
	 \asysrt(i) = (\afish,\alog[i])
	 \qquad
	 \alog[i] \sublog \alog[i]' \sublog \alog \qquad \alog[i] \subset \alog[i]'
	 }{
	 (\asysrt, \alog) \red[\tau] (\upd \asysrt i {(\afish, \alog[i]')}, \alog)
	 }{Prop}
\end{align*}

Rule \textsc{Local} describes the invocation of a command enabled at the $i$-th machine of the swarm $(\asysrt, \alog)$, assuming that machines are identified by their index; all the freshly generated events originate at the $i$-th machine and are non-deterministically merged in the global log by using the $\mergelog$ operator defined above. Finally, the swarm is updated by modifying the local log of the $i$-th machine with $\alog[i]'$, produced by the execution of the command $\acmd$.

Rule \textsc{Prop} defines event log propagation between machines.
The idea is to non-deterministically select a machine whose local log is a strict sublog of the global log, identify a larger sublog $\alog[i]'\sublog\alog$, and transfer events to the machine by assigning $\alog[i]'$ as its new local log.

Rules \textsc{Local} and \textsc{Prop} formalise respectively the of
effects command execution described \cref{sec:consumption}
and the non-deterministic log-shipping mechanism illustrated in \cref{sec:machine-interaction}.
Note that our formalisation acts on the logs which grow by appending newly generated events.
These features play an important role in the realisation of the local-first principle
and permit to formally represent\footnote{\cref{sec:full-example} demonstrates the subtleties of our model on our running example.}
the conflicts discussed in \cref{sec:in}.

Hereafter, $\ \weakred\ $ denotes a possibly empty sequence of
reduction steps and $\acmd$ ranges over commands; for swarm protocols we let
\[
 {\weakred}  = {\red^\star}
 \qqand[where] {\red} = {\red[\tau]} \ \ \cup \bigcup_{\acmd, \alogtype}\red[\afishact]
\]
and likewise for sequences of reduction steps of machines.
The following lemmata hold for coherent swarms.

\begin{restatable}[Coherence preservation]{lemma}{restateWfPreservation}\label{lem:wf-preservation}
  If $(\asysrt,\alog)$ is a coherent swarm and $(\asysrt,\alog)\red[\alpha](\asysrt',\alog')$ then $(\asysrt',\alog')$ is coherent.
\end{restatable}

\begin{restatable}[Eventual consistency]{lemma}{restateEvCons}\label{lem:ec}
  If $(\asysrt,\alog) = (\afish[1], \alog[1]) \parop \ldots \parop  (\afish[n], \alog[n]) \parop \alog$ is a coherent swarm,
  then $(\asysrt,\alog)\weakred(\afish[1], \alog) \parop \ldots \parop  (\afish[n], \alog) \parop \alog$.
\end{restatable}

Note how rules \rulename{Local} and \rulename{Prop} do not automatically guarantee consistency between machines and the sum of all information present in the swarm, i.e. the global log.
In our running example, a taxi machine might stay in its initial state while the rest of the swarm tracks the fulfilment of the passenger's transportation request, by virtue of not applying \textsc{Prop} for some time.
Addressing this issue is the purpose of the rest of this paper.

\subsection{Discussion of the computational model}\label{sec:discussion-model}

Similar to how a session type and its local projections are freshly instantiated for each session
our model always starts from the empty log.
The running example models the interactions for carrying out a single taxi ride,
it doesn't describe for example the ongoing series of rides each taxi participates in during one driver's shift.

Our computational model resembles the well-established {\em state machine replication} (\smr)~\cite{schneider1990implementing};
however it deviates considerably in the following aspects:

\noindent\emph{Machines are not identical:}
  we do  not necessarily consider all the machines in a swarm to be exact copies;
  indeed we expect them to be different to embody distinct roles in a coordinated interaction, as illustrated in~\cref{ex:local-bidding}.

\noindent\emph{Communication of effects instead of commands:}
  logs in \smr consist of sequences of commands, each of which needs to be executed in every single replica producing exactly the same outputs (determinism).
  We instead build logs of events that communicate the effects of command invocation in a particular machine.
  This implies that there is no expectation of reproducibility for the execution of a command.
  For instance, two different bidders in \cref{ex:local-bidding} may produce completely different offers when executing the command $\acmd[@][bid]$.
  For this reason, each command is executed just once on the originating machine and the produced effects are propagated to the remaining ones, which accept them as valid.
  In other words, our setting is that of collaborative interaction among trusted components.

\noindent\emph{Execution of commands:} in our model commands provide an
  interface to the environment in which swarms execute.
A command in a machine may be non-deterministically chosen by the
  environment and executed if enabled in the current state of the
  machine.
The execution of a command triggers events that allow machines
  to move to another state.

\noindent\emph{Machines may idle:}
  in contrast to session typing disciplines, a machine that enables commands is not obliged to execute one of them.
  For the progress of the swarm it is sufficient that one enabled command is eventually invoked.

\noindent\emph{Consistency among local logs:}
  while \smr ensures that of two sites' event logs one is always a prefix of the other, this is not true in our system as detailed in
  \cref{sec:consumption} (cf. \cref{fig:event-log-merging}).
  Our model only provides eventual consistency as per \cref{lem:ec}.
  
\noindent\emph{Stability vs Speculative computation:} in contrast to \smr, which advances local state only for the stable part of the event log, our model immediately computes the new states from sublogs as
  they are received from other machines.  This leads to intermediate states that can be invalidated by the later reception of a sublog that is \quo{merged into the past} as shown in \cref{sec:consumption} (cf. \cref{fig:event-log-merging}).

\section{Tool support}\label{sec:tooling}
Our theoretical development is accompanied by a set of software tools that support the implementation of swarms as a composition of type-checked TypeScript machines~\cite{typescript} and runs them based on the Actyx middleware~\cite{actyx}.
The ecosystem is depicted in \cref{fig:tools}.

\subsection{Execution of compiled machines}\label{sec:machine-runner}

\def\mkNode#1{\begin{tikzpicture}
  \useasboundingbox (-1ex,0)rectangle(1ex,0);
  \node[anchor=base,draw,fill=black,text=white,shape=circle,font=\footnotesize\bfseries,inner sep=1pt]{#1};
\end{tikzpicture}}

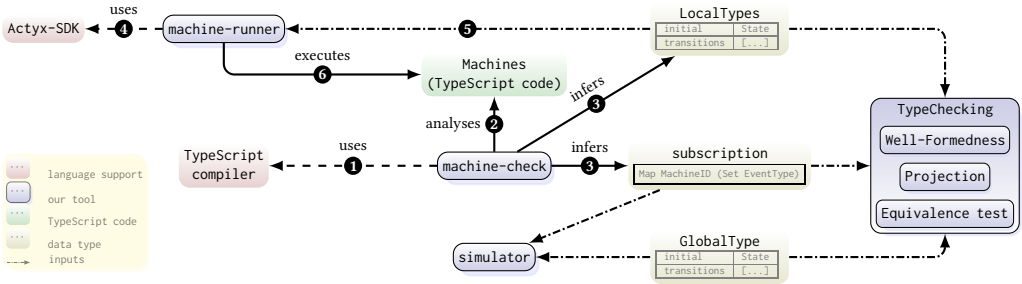
\begin{figure}[tb]
  \newcommand{\typerep}[1]{
  {#1}\\
  {\color{black!50}\scriptsize
	 \begin{tabular}{|l|l|}
		\hline
		initial & State
		\\\hline
		transitions & [...]
		\\\hline
	 \end{tabular}  
  }
}
\begin{tikzpicture}[node distance = 3cm,scale=.6, transform shape,
  every node/.style={
	 rectangle,
	 rounded corners,
	 minimum size=6mm,
	 align=center,
	 thin,
	 font=\ttfamily
  },
  every path/.style={
	 draw,
	 thick,
	 rounded corners,
	 -latex
  },
  used/.style={
	 top color=white,
	 anchor=center,
	 bottom color=red!50!black!20
  },
  dev/.style={
	 draw,
	 thin,
	 anchor=center,
	 drop shadow={color=blue!15, shadow scale = .95},
	 top color=white,
	 bottom color=blue!50!black!20,
	 font=\ttfamily
  },
  model/.style={
	 top color=white,
	 anchor=center,
	 bottom color=yellow!50!black!20,
	 font=\ttfamily
  },
  ts/.style={
	 top color=white,
	 anchor=center,
	 bottom color=green!50!black!20,
	 font=\ttfamily
  },
  lab/.style={
	 above=3pt,
	 font=\rmfamily
  },
  num/.style={
	anchor=center,
	fill=black,
	text=white,
	circle,
	minimum size=4mm,
	inner sep=0pt,
	font=\bfseries
  }
  ]
\node (sdk) [used] {Actyx-SDK};
  \node (runner) [dev] at(4,0) {machine-runner};
  \node (machines) [ts] at(10,-1) {Machines \\ (TypeScript code)};
  \node (lt) [model] at(15,0) {\typerep{LocalTypes}};
  \node (compiler) [used] at(4,-3) {TypeScript\\compiler};
  \node (check) [dev] at(10,-3) {machine-check};
  \node (subscription) [model] at(15,-3) {subscription\\
		\fbox{\color{black!50}\scriptsize Map MachineID (Set EventType)}
  };
  \node (simulator) [dev] at(10,-5) {simulator};
  \node (gt) [model] at(15,-5) {\typerep{GlobalType}};
  \node (typechecking) [dev, bottom color=white, top color=blue!50!black!20] at(20,-3) {
	 TypeChecking
	 \\[1ex]
	 \tikz\node[dev,draw]{Well-Formedness};
	 \\[1ex]
	 \tikz\node[dev,draw]{Projection};
	 \\[1ex]
	 \tikz\node[dev,draw]{Equivalence test};
  };
  \node (legend) [below= 4cm of sdk.west, anchor=west, fill=yellow!20, rounded corners, align=left, scale=.75, drop shadow={color=yellow!10, shadow scale = .95}, opacity=.5]{
		\tikz \node[used]{$\cdots$}; \quad \text{language support}
		\\
		\tikz \node[dev]{$\cdots$}; \quad \text{our tool}
		\\
		\tikz \node[ts]{$\cdots$}; \quad \text{TypeScript code}
		\\
		\tikz \node[model]{$\cdots$}; \quad \text{data type}
		\\
		\tikz \path[densely dashdotted] (0,0) -- (.7,0); \quad \text{inputs}
  };
\path[densely dashdotted] (gt.west) -- (simulator);
  \path[densely dashdotted] (gt.east) -| (typechecking.south);
  \path[densely dashdotted] (subscription) -- (simulator);
  \path[densely dashdotted] (subscription) -- (typechecking);
  \path[densely dashdotted] (lt) -| (typechecking.north);
  \path (check) edge[dashed] node[lab]{uses} node[num]{1} (compiler);
  \path (check) edge node[lab,above=0pt,left=5pt]{analyses} node[num]{2} (machines);
  \path (check) edge node[lab]{infers} node[num]{3} (subscription);
  \path (check) edge node[lab,sloped]{infers} node[num]{3} (lt);
  \path (runner) edge[dashed] node[lab]{uses} node[num]{4} (sdk);
  \path[densely dashdotted] (lt) -- node[num]{5} (runner);
  \path (runner) |- node[lab,pos=0.75]{executes} node[num,pos=0.75]{6} (machines);
\end{tikzpicture}
  \caption{Tool ecosystem}
  \label{fig:tools}
\end{figure}

The \texttt{machine-runner} library uses the Actyx SDK (cf. arrow \mkNode4 in \cref{fig:tools}) to drive machines written in TypeScript; more precisely, it employs \mkNode5 local types to interpret incoming events and execute \mkNode6 the corresponding machine logic.
The declaration of a machine revolves around the event types that it can handle.  Referring back to Listing~\ref{lst:ts-passenger}, we show the \texttt{Requested} event type on line 2 as an example.  Using a property called \texttt{type} to hold a string of singleton type (here: {\color{forestgreen}\texttt{'Requested'}}) is a customary way to express a discriminated (or \emph{tagged}) union in TypeScript, as shown on line 3.

Every machine state is represented by a class that derives from the \texttt{State} base class (or \emph{prototype} in JavaScript terms) provided by the \texttt{machine-runner} library.
This serves both as a marker for machine states and to carry the type parameter constraining all emitted events to a common type: the inherited \texttt{this.events} function used for example on line 28 is a utility for helping TypeScript to correctly capture the tuple type \texttt{[Selected, PassengerID]} (instead of the otherwise inferred array type \texttt{Events[]}) and assert that each of the event types conforms to type \texttt{Events}.

A program using the passenger's machine would start by constructing the initial state using for example \quo{\texttt{new InitialP('myID')}}.
Together with a suitable set of Actyx event tags (like {\color{forestgreen}\texttt{'taxiRide:12345'}} to denote this particular taxi ride protocol session) and a state change callback (see below) this initial state is then passed to the library's \texttt{runMachine} function.
This will set up a subscription for events with those tags using the Actyx SDK, where Actyx will first deliver all historic events already locally known and then switch to live mode.

Whenever an event is received, it is slated for consumption by the appropriate event handler method; to this end the handler method for the event's type is dynamically looked up in the JavaScript object underlying the state's class.
If that method takes only a single argument then the event is immediately consumed by calling the method, which returns the next state of the machine. Otherwise, the event is enqueued, awaiting the receipt of the following required event type etc. until the desired event sequence is complete and the method can be invoked with all arguments.
During this whole process, whenever the incoming event type does not have a matching handler or is not of the next required type in an argument list, the event is discarded as detailed in \cref{sec:consumption}.

Whenever the machine state changes (i.e. when $\stchange(\afish,\alog)$ computes a new value), the new state is passed to a function that the application passed to \texttt{runMachine} earlier---this scheme is termed a \emph{callback} in TypeScript (note that this language implements an imperative style with mutable bindings).
This could update a user interface or trigger an algorithm to compute reactions.
The state's command methods can therein be used to construct adequate event payloads for enabled commands, which would then be stored in Actyx using an SDK function and come back via the event subscriptions---now with metadata---to be applied to the current state and eventually trigger another invocation of the callback.

\subsection{Enforcing typing at run-time}\label{sec:machine-check}

Readers versed in TypeScript may have noticed that we glossed over a difficulty here: TypeScript types are fully erased at runtime, meaning that the \texttt{machine-runner} code will not be able to find the event handler method by using the event type, and it will also not know how many arguments that method takes and what its types are.
Therefore, the first responsibility of the \texttt{machine-check} build tool is to analyse \mkNode2 the TypeScript code and ascertain that all event types are declared such that they can be recognised at runtime on their \texttt{type} property---the handler method's name can then be constructed by prefixing the value of this property with {\color{forestgreen}\texttt{'on'}}.
The second responsibility is to extract the function signatures of all event handlers, check that each handler's name corresponds to the name of its first argument type, and then construct a per-state mapping from first event type to the list of following events (possibly empty).
This information is made available to \texttt{runMachine} by decorating~\cite{ts-decorator} the user-written state class: the \texttt{@proto} decorator transforms the class definition as it is loaded by the JavaScript VM, overriding the \texttt{reactions} method inherited from the \texttt{State} prototype.
In order to do that, the implementation of the \texttt{proto} function needs to have access to the \texttt{machine-check}'s extraction results.
This is done by importing \mkNode5 a source module generated by \texttt{machine-check} that contains all protocol information in JSON format~\cite{json}.

While the aforementioned duties of \texttt{machine-check} are crucial for \texttt{machine-runner}'s operation, the more interesting function of this build tool relates to the inference of local types and subscriptions (arrows \mkNode3 in \cref{fig:tools}) as well as initiating the type-checking process on swarm protocols.
To this end, the TypeScript compiler is used \mkNode1 as a library to obtain \mkNode2 a fully typed AST representation of the user program.
Since we are only interested in machines, our entry points are \texttt{State} subclasses that are marked as initial states by a documentation comment starting with \quo{Initial state for role}, as is shown on line 5 of Listing~\ref{lst:ts-passenger}.
This comment serves the secondary purpose of naming the role this machine aspires to play (P for passenger in this example).
The \texttt{@proto} decorator on line 6 not only has its runtime duties as explained above, it also carries in its argument the name of the swarm protocol that provides the context for the role name---we discuss both concepts in detail in the following sections;
\texttt{machine-check} expects to find the definition of the swarm protocol in a correspondingly named file in JSON format.
Finally, \texttt{machine-check} assembles the lists of command and event handler methods by inspecting (arrow \textcircled{5}) a state class's method names and signatures and follows up with recursively processing the result types of event handlers in the same fashion.
Any event type seen in an event handler argument list is automatically added to the subscription set of the machine (needed for the projection as explained in \cref{sec:proj}).

\subsection{Type-checking, simulation and more}

As a result of the analysis described above \texttt{machine-check} has assembled the following pieces for each machine definition within the user program: swarm protocol, role name, subscriptions, states, and transitions.
Each such tuple is then passed---again in JSON format---to the \texttt{typechecking} tool, our third artifact contribution, written in Haskell.
This tool first checks that the provided swarm protocol and subscription are well-formed (according to the rules presented in \cref{sec:wf}), computes the projection for the given role, and finally checks the inferred machine type for equivalence to the projection result (where state names are immaterial).

In addition, we provide an auxiliary tool written in Haskell for
the simulation of the formal operational semantics of the model. For a given
protocol and subscription, the tool computes the projections and simulates the execution of
swarms consisting of machines according to those projections. It supports both exhaustive and random
generation of traces up to a given length. The tool has been used for
checking claims and results about our running example.

The aforementioned tools are detailed in an artifact submission accompanying
this paper. Through an example project, the accompanying paper also
demonstrates the use of the inferred machine type to generically render a
machine UI. Besides showing the current state of the computation, the UI gives
the user the possibility to interact with machines by invoking enabled
commands, where command arguments are gathered using automatically generated
HTML forms.

\subsection{Noteworthy properties of Actyx}\label{sec:actyx}

The Actyx middleware provides a reliable durable pub--sub mechanism for tagged event logs as well as a coordination-free total order of all events, as we formalise in \cref{sec:machine-interaction}.
Actyx is a peer-to-peer system built from \emph{nodes} running on network devices.
The event order is realised by combining a Lamport clock~\cite{lamport.clock} for causality-preserving logical timestamps (additionally incremented for each emitted event) with a unique node identifier based on the public key of an ED25519 pair.
Each node may host multiple machines whose event logs are separated by using distinct tags; each event carries a set of tags, and event logs are queried by a boolean combination of tags that correspond to union/intersection of the thus tagged sets of events.
As Actyx nodes are merely a deployment unit our model deals only in machines---nodes don't add semantic differences.

The sublog relation (\cref{def:sublog}) is owed to how local logs are characterised in Actyx.
Events are written to a per-node single-writer append-only log slice.
Local logs are characterised by the locally available length of each slice.
This yields a much more concise representation than marking the presence of individual events, facilitating periodic broadcast to stimulate reliable replication.

There are two modes of live event subscriptions: the \texttt{subscribe} SDK method yields a stream that will---after emitting all past events---emit each fresh event as soon as it is received or locally emitted.
The \texttt{subscribeMonotonic} SDK method does the same thing, but only up to the point where a fresh event would need to be sorted earlier than the previously emitted event (which is termed \emph{time travel} in Actyx).
\texttt{machine-runner} uses the latter function and restarts the computation from the initial state whenever time travel occurs; note that the historical part of the emitted event log will now contain the event that caused the time travel.

\section{Swarm protocols}\label{sec:gt}

A \emph{swarm protocol} (hereafter also called \emph{protocol} for short) describes the intended overarching event log structure realised by a swarm of machines;
it corresponds to a global type in the terminology of session types, with our machines playing a similar role to local types.
The protocol captures the overall communication structure as well as the details relevant for implementing it with machines.
As it is customary with behavioural types, swarm protocols rely on an idealised environment where all communication is infallible and instantaneous.
The link to the realisation in terms of machines is given in the following section by way of a projection operation.

To simplify the technical treatment of recursive types we define a \emph{swarm protocol} as a \emph{regular term}\footnote{
  A term is regular if it consists of finitely many \emph{distinct} subterms. This condition ensures that the language generated by the co-inductive grammar is finitely representable either using the so-called \quo{$\mu$ notation}~\cite{Pierce02} or as solutions of finite sets of equations~\cite{Courcelle83}. Also, there is a correspondence between these structures and finite-state automata; the interested reader is referred to \cref{sec:regtrees2lts} or, for a comprehensive treatment to~\cite{Courcelle83}.} derivable from the co-inductive grammar
\[
 \agt \stackrel{\text{co}}\bnfdef \agsum
\]
where $\arole[]$, $\arole[1]$, $\arole[2]$, etc. range over roles.
We write $\agt=\zero$ when $I$ is empty and $\bigl(\sum_{i \in I}\gsumprefix\bigr).\agt$ for a common continuation,
that is when $\agt_i = \agt$ for all $i \in I$.
Intuitively, the protocol progresses by some role $\arole[i]$ invoking command $\acmd[i]$, appending a non-empty event sequence of type $\alogtype[i]$ to the global log and continuing as protocol $\agt[i]$.
As discussed at the end of \cref{sec:formalisation}, the resolution of the choice specified in a swarm protocol is not coordinated among the instances of the roles involved in the choice (i.e.\,there is no unique selector).
In fact, instances of different roles involved in the choice may enable commands at the same time as well as different instances of the same role may enable different commands (recall that each machine tracks a separate local log and event replication is asynchronous).
This is in contrast to most other behavioural type systems hitherto, which do not permit such race conditions.

\begin{definition}[Determinism]\label{def:gt-det}
  A swarm protocol $\arec[{\agt}][{\agsum}]$ is \emph{log-deterministic} if the event types ${\alogtype[i]}[0]$ are pairwise different and all $\agt[i]$ are log-deterministic.
  \(\agt\) is \emph{command-deterministic} if the tuples $(\acmd[i],\arole[i])$ are pairwise different and all $\agt[i]$ are command-deterministic.
  \(\agt\) is \emph{deterministic} if it is log-deterministic and command-deterministic.
\end{definition}

Hereafter we only consider deterministic swarm protocols.
Note that determinism is evidently decidable on swarm protocols due to the regularity constraint.

\begin{example}[Taxi service]\label{ex:taxi-service}
  We can formalise the protocol in \cref{ex:local-bidding} as follows:
  \begin{align*}
    \agt =&\
    \gsumprefix[][Request][Requested][P]\gtpref\gsumprefix[][Offer][Bid,BidderID][T]\gtpref\agt[\text{auction}]
    \\
    \agt[\text{auction}] =&\
    \gsumprefix[][Offer][Bid,BidderID][T]\gtpref\agt[\text{auction}] +
    \gsumprefix[][Select][Selected,PassengerID][P]\gtpref\agt[\text{choose}]
    \\
    \agt[\text{choose}] =&\
    \gsumprefix[][Arrive][Arrived][T]\gtpref\gsumprefix[][Start][Started][P]\gtpref\agt[\text{ride}] +  \\&\
    \gsumprefix[][Cancel][Cancelled][P]\gtpref\gsumprefix[][Receipt][Receipt][O]\gtpref\zero
    \\
    \agt[\text{ride}] =&\
    \gsumprefix[][Record][Path][T]\gtpref\agt[\text{ride}]
     + \gsumprefix[][Finish][Finished,Rating][P]\gtpref\gsumprefix[][Receipt][Receipt][O]\gtpref\zero
  \end{align*}
The structure in terms of commands and roles is straightforwardly induced from \cref{ex:local-bidding},
  with event log types filled in according to more specific requirements.
The event type $\aeventtype[][BidderID]$ represents identifying information illustrating that not all events are of interest to all roles: it is reasonable to assume that the office does not need to know all bidders, it only needs to know which taxi was $\aeventtype[][Selected]$.
It is  straightforward to check that $\agt$ is both log- and command-deterministic.
  \finex
\end{example}

Mirroring the formulation of machines we ascribe operational semantics to a swarm protocol via the generation and processing of an event log.
The main difference in state computation is that swarm protocols generate and consume logs instead of events,
as illustrated with the $\acmd[][Offer]$, $\acmd[][select]$, and $\acmd[][Finish]$ commands in the example above.

For a swarm protocol $\agt=\agsum$ we write $\agt\red[\afishact]\agt'$ if
there is $j \in I$ such that $\acmd[] = \acmd[j]$,
$\alogtype[] = \alogtype[j]$, and $\agt' = \agt[j]$.
The state reached by a swarm protocol $\agt$ after processing a log
$\alog$ is
computed using an extension of the transition function $\stchange$ on machines, namely as follows:
\begin{align*}
  \Stchange(\agt,\emptylog) &= \agt
  &
  \Stchange(\agt,\alog) &=
    \begin{cases}
      \Stchange(\agt', \alog'')
      &
        \text{if }\ \agt \red[\afishact] \agt'\text{ and }\eventtyping[][\alog'][\alogtype]\text{ and }\alog=\alog' \logcat \alog''
         \\
             \bot
       &
        \text{otherwise}.
     \end{cases}
\end{align*}

Analogously to the definition for machines, the transition function $\Stchange(\agt,\alog)$ returns the continuation of the swarm protocol $\agt$ after processing the entire log $\alog$.
We stress that $\Stchange$ is a partial function on swarm protocols,
in fact it is undefined when log $\alog$ cannot be generated according to $\agt$
(unlike the definition of $\stchange$ on machines, which is a total function since it just discharges unrecognised events).
Note also that $\Stchange$ is well-defined over log-deterministic
swarm protocols because a log in a branch cannot appear as a prefix of
the logs of the remaining branches of the choice.

The operational semantics of a swarm protocol are defined as a labelled transition system given by the following induction rule:
\begin{align*}
  \mathrule{
  \Stchange(\agt, \alog) \red[\afishact] \agt'
  \qquad
  \eventtyping[][\alog'][\alogtype]
  \qquad \alog'\ \text{ log of fresh events}
  }{
  (\agt,\alog) \red[\afishact] (\agt,\alog \logcat \alog')
  }{G-Cmd}
\end{align*}
A swarm protocol $\agt$ with log $\alog$ enables command $\acmd$, upon whose invocation the log is extended with fresh events $\alog'$ of type $\alogtype$ before possibly allowing another command to be invoked.
The freshness of the events in $\alog'$ can e.g.\ be guaranteed by the inclusion of node ID and logical timestamp.

\begin{example}[Idealised taxi service]
Consider the protocol from \cref{ex:taxi-service}, starting out with $(\agt,\emptylog)$.
After invoking the $\acmd[][Request]$ command our log contains $\aevent[][Requested]$ and we reach state $\agt[\text{Auction}]$.
Two bids later the passenger makes their selection, leading us to
  \[\Stchange(\agt,\mklog{\aevent[][Requested],\aevent[A][Bid], \aevent[A][BidderID],\aevent[B][Bid], \aevent[B][BidderID], \aevent[][Selected], \aevent[][PassengerID]})=\agt[\text{choose}]\]
  with $\agt[\text{choose}]$ offering two options: either the passenger invokes $\acmd[][Cancel]$ or the taxi $\acmd[][Arrive]$s.
\finex
\end{example}

\section{Projection}\label{sec:proj}
For the definition of our projection operation it is convenient to introduce a textual presentation of machines equivalent to the automata-based presentation used so far. Let $\acmdrel$ denote a finite function mapping commands to non-empty log types; we allow ourselves to treat $\acmdrel$ as set (the graph of function $\acmdrel$) and e.g. write $\acmd \cte \alogtype \in \acmdrel$ for $\acmdrel(\acmd) = \alogtype$ or else write $\{\afishact[1], \ldots, \afishact[h]\}$ for the function $\acmdrel$ mapping $\acmd[i]$ to $\alogtype[i]$ for each $i \in \{1, \ldots h\}$.
$\aeventtype[i]$ ranges over event types.

Similarly to swarm protocols, the textual presentation of our machines is a regular term\footnote{
  The correspondence between these regular terms and finite-state automata yields exactly the presentation of machines in terms of finite-state automata that we have adopted so far. 
} of the following co-inductive grammar:
\begin{align}\label{eq:lt}
  \afish \stackrel{\text{co}}\bnfdef
  \asumnew[@][{\areaction[1], \cdots, \areaction[n]}]
\end{align}
and we abbreviate $\asumnew[@][{\areaction[1], \cdots, \areaction[n]}]$ as $\acmdrel \cdot \zero$ when $n = 0$ and as $\asumnew[][{\areaction[1], \cdots, \areaction[n]}]$ when $\acmdrel$ is the empty map. We also write ${\branchsep_{1 \leq i \leq n}\inp[i][{\alogtype[]}]\afish[i]}$ in place of $\asumnew[][{\areaction[1], \cdots, \areaction[n]}]$.

In turning our attention to projection operations we first note that the responsibility for driving the protocol forward is distributed across the participants: each transition in the swarm protocol is labelled with one role that may trigger it by invoking the command.
Each machine plays one role, whose machine specification is obtained by the projection operation $\agt\aproj[{\arole[]}][]$.
Note that multiple machines may implement the same role.

One could naively define $\agt\aproj[@][]$ such that each transition in $\agt$ produces a series of event transitions in the machine plus a command invocation on the originating state if the role matches.
\[
  \Bigl(\agsum\Bigr)\aproj[@][]=
  \asumnew[@]
  [{\branchsep_{i\in I}\inp[i][{\alogtype[]}]\agt[i]\aproj[{\arole[]}][]}] \qqand[where] \acmdrel = \{(\acmd[i]\cte\alogtype[i]) \sst \arole[i]=\arole[] \text{ and } i \in I\}
\]
Albeit simple, this projection scheme generates unnecessarily large machines in all but the most trivial cases.
More crucially, forcing each machine to process all events is undesirable for reasons of security and efficiency.
It would be highly desirable to allow some information to be kept secret from certain roles
(like $\aeventtype[][passengerID]$ in \cref{ex:taxi-service}),
and it would be most efficient if every role processed just enough information to correctly enable and disable command invocations.
We therefore define a more appealing construction.

Our projections are based on the notion of whether a machine shall process a certain type of event.
Formally, the projection operation is parameterised by a \emph{subscription}, namely a map $\reactsto$ assigning to each role the set of event types that it reacts to.
Given a set of log types $E$, let $\filter[{\_,E}]$ be a function transforming a log type, retaining only the event types in $E$ while preserving their relative order.
Intuitively, subscriptions correspond to topics in a publish--subscribe framework whereby processes declare which kinds of messages they are interested in receiving.

\begin{definition}[Projection]\label{def:proj}
  Given a swarm protocol $\agt$ and a subscription $\reactsto$, the
  \emph{projection of $\agt$ over a role $\arole[]$ with respect to
    $\reactsto$}, written $\agt\aproj$, is defined as
  follows:
  \begin{align*}
    \Bigl(\agsum\Bigr)\aproj =
    \asumnew[{
      \{\acmd[i]\cte\alogtype[i] \sst \arole = \arole[] \text{ and } i \in I\}
      }][\displaystyle{\branchsep_{j \in J}}{\filter[{\alogtype[j], \reactsto[{\arole[]}]}]\, {\colorSymb ?}\ (\agt[j]\aproj)}]
  \end{align*}
  where $J = \{ i \in I \sst \filter[{\alogtype[i], \reactsto[{\arole[]}]}] \neq \emptylog\}$.
\end{definition}

Notice that we omit the projection of a branch when a role $\arole[]$
is not subscribed to any of the event types emitted by the command that selects that branch.
We opted for this simplification of the formalism because our well-formedness conditions (cf. \cref{sec:wf}) ensure that if a role is involved in the continuation it will subscribe to the the first event in the branch.
Further, note that this pruning applies to branches in isolation, later states reachable by other paths remain part of the projection.

\begin{example}\label{ex:proj-well-formed}
Let \(\reactsto=\bigl\{
    \arole[][P]\mapsto E,
    \arole[][T]\mapsto E,
    \arole[][O]\mapsto E\setminus\{\aeventtype[][BidderID]$, $\aeventtype[][PassengerID] \}
  \bigr\}\)
  with $E$ being the set of all event types in the protocol $\agt$ defined in \cref{ex:taxi-service}.
The projections of $\agt$ over the different roles $\agt\aproj[{\arole[][P]}]  =  \afish[][P]$, $\agt\aproj[{\arole[][T]}]  =  \afish[][T]$ and  $\agt\aproj[{\arole[][O]}]  =  \afish[][O]$ are in \cref{fig:taxi}.
\begin{figure}[t]
\begin{tikzpicture}[gt, scale = .8, node distance = 4mm and 18mm, font=\footnotesize]
      \node[initial] (p) {$\arole[][P]$};
		\node[state,right = of p] (p') {};
		\node[state,right = 1cm of p'] (p'') {};
		\node[state,right = of p''] (1) {};\node[state,right = of 1] (1') {};
		\node[state,below = of 1] (1'') {};
      \node[state,right = of 1'] (2) {};\node[state, above right = of 2] (3) {};\node[state,right = of 3] (4) {};\node[state,below right = of 4] (4') {};
      \node[state, below = of 4'] (0) {};\path (p) edge[loop below] node[xshift=8mm]{${\afishact[][Request][Requested]}$} ();
      \path (p) edge node{${\inp[][Requested]}$} (p');
      \path (p') edge node{${\inp[][Bid]}$} (p'');
      \path (p'') edge node{${\inp[][BidderId]}$} (1);
      \path (1) edge[loop above] node{${\afishact[][Select][\mklogtype{Selected,PassengerID}]}$} ();
      \path (1) edge[bend left] node{${\inp[][Bid]}$} (1'');
      \path (1'') edge[bend left] node{${\inp[][BidderId]}$} (1);
      \path (1) edge node{$\inp[][Selected]$} (1');
      \path (1') edge node[sloped]{$\inp[][PassengerID]$} (2);
      \path (2) edge[loop above] node{${\afishact[][Cancel][Cancelled]}$} ();
      \path (2) edge[bend right=20] node[below]{$\inp[][Cancelled]$} (4');
      \path (2) edge node[below,sloped,near end]{$\inp[][Arrived]$} (3);
      \path (3) edge node[sloped]{$\inp[][Started]$} (4);
      \path (3) edge[loop above] node{${\afishact[][Start][Started]}$} ();
      \path (4) edge[loop above] node{${\afishact[][Finish][Finished]}$} ();
      \path (4) edge[loop below] node{${\inp[][Path]}$} ();
      \path (4) edge node[sloped]{$\inp[][Finished]$} (4');
      \path (4') edge node{$\inp[][Receipt]$} (0);
    \end{tikzpicture}
	 \\
	 \begin{tikzpicture}[gt, scale = .8, node distance = 4mm and 20mm, font=\footnotesize]
      \node[initial] (t) {$\arole[][T]$};
		\node[state,right = of t] (1) {};\node[state,right = of 1] (1') {};
		\node[state,below = of 1] (1'') {};
      \node[state,right = of 1'] (2) {};\node[state,above right = of 2] (2') {};
      \node[state,right = of 2'] (3) {};\node[state,below right= of 3] (3') {};
		\node[state, below=  of 3'] (0) {};\path (t) edge node{${\inp[][Requested]}$} (1);
      \path (1) edge[loop above] node{${\afishact[][Offer][\mklogtype{Bid,BidderID}]}$} ();
      \path (1) edge[bend left] node{${\inp[][Bid]}$} (1'');
      \path (1'') edge[bend left] node{${\inp[][BidderId]}$} (1);
      \path (1) edge node{$\inp[][Selected]$} (1');
      \path (1') edge node[sloped]{$\inp[][PassengerID]$} (2);
      \path (2) edge[loop above] node{${\afishact[][Arrive][Arrived]}$} ();
      \path (2) edge[below] node[sloped, near end]{$\inp[][Arrived]$} (2');
      \path (2') edge node[sloped]{$\inp[][Started]$} (3);
\path (3) edge[loop above] node{${\afishact[][Record][Path]}$} ();
      \path (3) edge[loop below] node{${\inp[][Path]}$} ();
      \path (3) edge node[sloped]{${\inp[][Finished]}$} (3');
      \path (2) edge[bend right=20] node[below]{${\inp[][Cancelled]}$} (3');
      \path (3') edge node{$\inp[][Receipt]$} (0);
    \end{tikzpicture}
	 \\
	 \begin{tikzpicture}[gt, scale = .8, node distance = 4mm and 20mm, font=\footnotesize]
      \node[initial] (o) {$\arole[][O]$};
		\node[state,right = of o] (o'') {};
		\node[state,right = of o''] (o') {};\node[state,right = of o'] (1) {};\node[state,right = of 1] (1') {};
      \node[state,right = of 1'] (1'') {};\node[state,right = of 1''] (2) {};\node[state,below = of 2] (0) {};\path (o) edge node{${\inp[][Requested]}$} (o'');
      \path (o'') edge node{${\inp[][Bid]}$} (o');
      \path (o') edge node{${\inp[][Selected]}$} (1);
      \path (o') edge[loop above] node{${\inp[][Bid]}$} ();
      \path (1) edge node{$\inp[][Arrived]$} (1');
      \path (1') edge node{$\inp[][Started]$} (1'');
      \path (1'') edge node{$\inp[][Finished]$} (2);
      \path (1'') edge[loop above] node{${\inp[][Path]}$} ();
      \path (2) edge[loop above] node{${\afishact[][Receipt][Receipt]}$} ();
      \path (2) edge node{$\inp[][Receipt]$} (0);
      \path (1) edge[sloped, bend right=20] node[below]{${\inp[][Cancelled]}$} (2);
    \end{tikzpicture}
\caption{Projections of the running example as automata (see \cref{sec:full-example} for the textual presentation)}
   \label{fig:taxi}
 \end{figure}
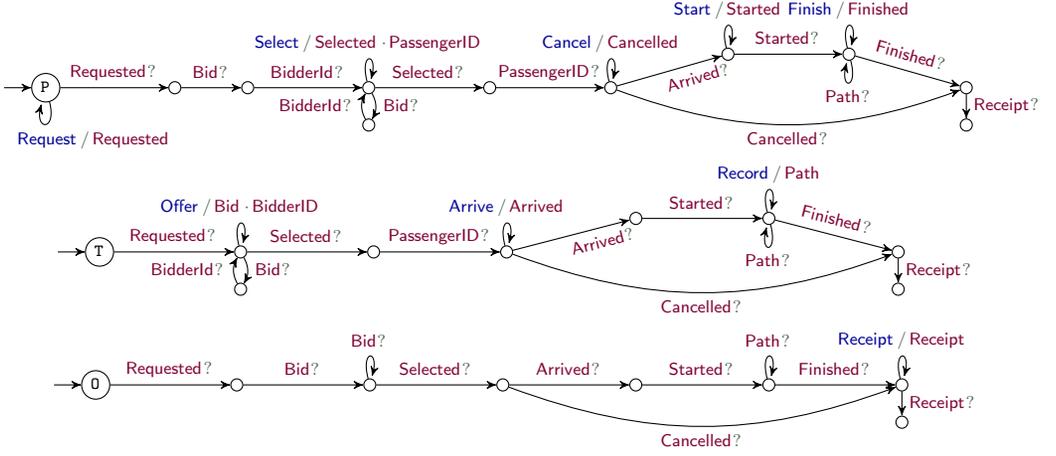

\finex
\end{example}

\section{Well-formedness}\label{sec:wf}

We now  focus on the \emph{well-formedness} conditions of our swarm protocols.
As is standard in behavioural types, sufficient conditions are
established on global specifications that guarantee relevant
properties on projections such as deadlock or lock freedom and
absence of orphan messages.
Noticeably, the properties of interest to us are quite different from
those common in standard settings (cf.~\cref{sec:deadlockfree} for a discussion of those).
Instead, we aim to guarantee that eventual consensus is reached even when some of the participants
make choices that are discordant due to their incomplete view on the global log.
The idea is that transitory deviations are \emph{tolerated} provided
that consistency is eventually recovered, which happens once information has
sufficiently spread within the swarm.
For instance, a taxi in our running example may keep bidding for a passenger's auction
after the passenger has made their selection as long as the selection event has not yet been received.
This temporary inconsistency is recognised and resolved once the events
have propagated to the deviating taxi and the passenger, respectively.

Realising swarms with this property is not straightforward.
The rest of this section illustrates the problems arising in our
setting with a few examples.
For each problem we identify sufficient conditions on our swarm protocols
that rule out the problem (\cref{sec:full-example,sec:problems,sec:anomalies} show coordination issues in realistic scenarios based on our running example).
These conditions culminate in our definition of \emph{well-formedness}
(cf. \cref{def:wf}).

\subsection{On causality and propagation}\label{sec:causpropprob}
The first problem we look at is related to how a command is disabled once it has been invoked.
In our setting, this boils down to fine tuning the registration of roles to event types.
For example, if  a command $\acmd$ should be enabled only after another, say $\acmd'$ has been executed, then the role executing $\acmd'$ should be subscribed to some event type emitted in response to the execution of $\acmd$.
Another example is that a command can stay perpetually enabled if the role executing it is oblivious of all resulting events.
An instance of these issues is given in \cref{ex:causal-dep} in \cref{sec:problems} for our running example.

Another class of problems is caused by the fact that events propagate asynchronously within a swarm and that an emission of multiple events is not guaranteed to reach all other machines as one atomic transmission as illustrated by \cref{ex:bad-propagation}.
This anomaly may be excluded by a runtime system that never applies the \rulename{Prop} rule to a strict subset of the event log emitted by a single command, i.e. it treats the log from each command invocation as an atomic unit.
We chose to not restrict the way in which a runtime system should propagate events between network sites because we consider it important that implementations be free to optimise their strategy in different ways (e.g. for latency, bandwidth, efficiency, or consistency).

We define the \emph{active} roles of a swarm protocol $\agt$ as those that can select one of the branches in the top-level choice of $\agt$ and---given a subscription $\reactsto$---the \emph{roles} of $\agt$ as those that can invoke commands or are subscribed to events in  $\agt$.
Formally,
\begin{align*}
  \activer\Big(\agsum\Big) & = \{\arole[i] \sst i\in I\}
  \\
  \roles\Big(\agsum\,,\reactsto\Big) & =
   \bigcup_{i\in I} \ \big(
   \{\arole[] \in \dom \reactsto \sst \arole[i] = \arole[]\ \lor\ {\alogtype[i] \cap \reactsto(\arole[]) \neq \emptyset} \}
   \ \cup\
   \cmpts[{\aspecdef[{\agt[i]}]}]\big)
\end{align*}
(note that the latter is a coinductive definition).
With this notation we define the following sufficient condition for avoiding the aforementioned problems.

\begin{definition}[Causal consistency]\label{def:cc}
  A swarm protocol $\agsum$ is \emph{causal-consistent in a subscription $\reactsto$} if for all $i \in I$
  \begin{enumerate}
  \item\label{it:exec}
	 $\alogtype[i] \cap \reactsto[\arole] \neq \emptyset$, and
  \item\label{it:active} $\arole[] \in \activer[{\agt[i]}]$ implies
	 $\alogtype[i] \cap \reactsto[{\arole[]}] \neq \emptyset$ and for
	 all $\arole[]' \in \roles[{\aspecdef[{\agt[i]}]}]$,
	 $\alogtype[i] \cap \reactsto[{\arole[]'}] \subseteq \alogtype[i]
	 \cap \reactsto[{\arole[]}]$
  \end{enumerate}
\end{definition}
Condition~\eqref{it:exec} requires that the role that performs one of
the commands $\acmd[i]$ should observe some of the corresponding
emitted events $\alogtype[i]$.
This simple mechanism ensures that repeated command invocation can only occur where foreseen in the swarm protocol.
Condition~\eqref{it:active} ensures the adequate tracking of causality for subsequent command invocations.
The first part ensures that the immediately following command must wait for the enabling transition to occur,
while the second part guarantees the ordering of the subsequent command's generated events
after all events from the preceding command that are observed by some role in the further evolution of the protocol.

\begin{example}\label{ex:running-is-causal-consistent}
The protocol of the running example  $\agt$ in \cref{ex:taxi-service} is causal-consistent for the subscription $\reactsto$
   in \cref{ex:proj-well-formed}. In fact, the commands generate logs that start
  with pairwise-different event types. Hence, the conditions  straightforwardly hold for roles $\arole[][P]$ and $\arole[][T]$, which
  observe every event. For  $\arole[][O]$, we observe that they only execute the command $\acmd[][receipt]$; which should be performed after
  $\aeventtype[][Cancelled]$ or $\aeventtype[][Finished]$, which are also observed by $\arole[][O]$.
  \finex
\end{example}

\subsection{On distributed choices}\label{sec:choiceprobs}

The next anomaly we study is caused by the fact that our model permits multiple roles to be active at the same time without coordination---this property is essential for perfect availability as demanded by local-first cooperation.
Such behaviour would be ruled out in all the global type systems we are aware of.
Our strategy for coping with the inevitably arising conflicts is that we permit machines to make inconsistent local decisions but reconcile those once the corresponding events have propagated to all relevant parties; \cref{ex:bad-choice} in \cref{sec:problems} illustrates how the accounting office in our example can make a choice which is inconsistent with the decisions taken by other participants.
We fix this by requiring \emph{determinacy}.
\begin{definition}[Determinacy]\label{def:cd}
  A swarm protocol $\agt=\agsum$ is \emph{determinate with subscription $\reactsto$} if it is causal-consistent and
  $\arole[]\in\roles[{\agt[i],\reactsto}]$ implies $\alogtype[i]{[0]}\in\reactsto(\arole[])$ for all $i\in I$.
\end{definition}

This definition of determinacy is prompted by our determinism rule \cref{def:gt-det}: we identify a branch by the first event type of its emitted log.
Note that a role involved in one branch but not in another may invoke commands that are later invalidated
without that role being able to recognise this situation;
we will explore mechanisms for compensating such errors in future work, which may require strengthening the rule above.

We note that in conjunction with determinacy condition~\eqref{it:active} in \cref{def:cc} can be relaxed
(by allowing roles to skip some events that occur in between the first one and some agreed last event),
but this would make the constructions and the proofs more convoluted.
\begin{example}\label{ex:running-is-determinate}
  The swarm protocol $\agt$ in \cref{ex:taxi-service} is  determinate for
  the subscription $\reactsto$ in \cref{ex:proj-well-formed}.
\finex
\end{example}

\subsection{On interference}\label{sec:interference}

Events emitted by the losing parties to a conflict should be ignored in order to let every machine eventually agree on each choice.
Each machine must locally be able to ignore such events to avoid the last class of problems described below.

Consider a variation of our taxi ride protocol that reuses the $\aeventtype[][finished]$ event type for cancellation.
Due to the race condition between cancellation and $\aeventtype[][arrived]$ marking the ride's begin, the cancellation may actually be mistaken for the ride being immediately finished if the participant does not observe the $\aeventtype[][arrived]$ event.
We avoid this confusion by requiring that any branch of a swarm protocol is communicated using a
dedicated event type, i.e. that event type cannot be emitted by any other command.
We formulate this notion in terms of the set
$\subterms(\agt)$ of all subterms (incl. indirect) of a swarm protocol $\agt$.
Recall that this set is finite because our swarm protocols are regular.

In what follows we write $\events[\agt]$ and $\rdy[\agt]$ respectively
for the sets of all event types and the ones that identify branches;
formally, if $\agt = \agsum$ then
\[
\events[\agt] = \bigcup_{i\in I}\left({\events[{\agt[i]}] \cup \bigcup_{j}\alogtype[i]}[j]\right)
\qqand
\rdy[\agt] = \bigcup_{i\in I}{(\{\hd{{\alogtype[i]}}\} \cup \rdy[{\agt[i]}])}
\]
(observe that $\events[\zero] = \rdy[\zero] = \emptyset$ and that we define these sets coinductively, meaning the greatest fixpoint; recall that swarm protocols are regular trees, wherefore these are finite sets computed in finite steps).

A swarm protocol $\agt$ is \emph{invariant under event type $\aeventtype$}
if either
(i) $\aeventtype$ does not appear in $\agt$, i.e. $\aeventtype\not\in\events[\agt]$ or
(ii) it only appears as part of the same choice, i.e. there is a unique $\agt'\in\subterms(\agt)$
such that $\agt'\red[{\afishact[][\acmd][\alogtype]}]$ and $\aeventtype \in \alogtype$.

\begin{definition}[Confusion-freeness]\label{def:wb}
  A swarm protocol $\agt$ is \emph{confusion-free} if $\agt$ is invariant
  for all event types in $\rdy[\agt]$.
\end{definition}

\begin{example}\label{ex:running-confusion-free}
   It is easy to check that the protocol of the running example  $\agt$ in \cref{ex:taxi-service} is invariant for all
   types. The only  type appearing in two guards is $\aeventtype[][Receipt]$; however, the occurrences are
   associated to the same subterm.
   Hence, the protocol is confusion-free.
  \finex
\end{example}

\subsection{Putting the pieces together}\label{sec:pieces}
With this, we can finally state our \emph{well-formedness} condition.

\begin{definition}[Well-formedness]\label{def:wf}
  A swarm protocol $\agt=\agsum$ is \emph{well-formed with respect to a subscription $\reactsto$}
  (\wf\ for short) if
  \begin{enumerate}
  \item ${\agt}$ is causal-consistent, determinate, and confusion-free; and
  \item $\agt[i]$ is \wf\ for all $i \in I$;
  \end{enumerate}
\end{definition}

Note that well-formedness is defined co-inductively and
decidable on swarm protocols.

\begin{example}\label{ex:well-formed}
\cref{ex:running-is-causal-consistent,ex:running-is-determinate,ex:running-confusion-free} imply that
the protocol  $\agt$ in \cref{ex:taxi-service}
is well-formed with respect to the subscription
$\reactsto$ given in \cref{ex:proj-well-formed}.
\finex
\end{example}

The following result ensures that projection preserves determinism in well-formed protocols.

\begin{restatable}{proposition}{restateProjdet}
Let $\agt$ and $\reactsto$ respectively be a swarm protocol and a subscription.
If $\agt$ is \wf\
then $\agt\aproj$ is deterministic for all $\arole[]$.
\end{restatable}

Well-formed swarm protocols guarantee that the local machines reach eventual consensus~\cite{8919675} on each choice, as we show next.
However, anomalies\footnote{Examples, not necessary for the understanding of
the paper, are in \cref{sec:anomalies} for the reviewers'
convenience.} occur at the level of systems:
\begin{itemize}
\item Machines could have commands enabled that would be disabled if
  the model were synchronous; this may lead to the emission of events
  that need to be ignored later.
\item Events are ignored according to their type only, therefore even
  after full propagation of the events in the global log a machine may
  process events stemming from the anomalous invocation of a command.
\end{itemize}
We note that the first anomaly above is inherent to local-first architecture requirements.

The second anomaly can be avoided by a runtime system that tracks full causality information.
We chose to not require full causality tracking since it imposes additional storage, communication, and computation requirements on the implementation.
Our weaker causality model supports deployment on less capable hardware where needed.

\section{Correct Realisations of swarm protocols}\label{sec:correct}
We now turn our attention to the formal characterisation of
{\em correct implementations} of  swarm protocols. As discussed in the previous
sections, we  deviate from the usual expected  properties of  mainstream
(multiparty) session types, such as communication safety, session fidelity, and progress (i.e.,
absence of  deadlocks or its variants).
We first note that there are no communication mismatches in our model because every machine
simply ignores unexpected or unwanted events (recall the definition of $\stchange$ in
\cref{sec:machine-single}).
Session fidelity instead advocates  implementations that behave as described by their types,  which
customarily means that the states of all components are always aligned with the
global state of the protocol.
Contrastingly, we aim to tolerate deviations provided that all
machines eventually agree on the state of the execution of the protocol.
In our setting, an implementation may be correct even if machines temporarily diverge,
executing different branches of the protocol; this is quite expected if we allow independent
decisions taken based on incomplete views of the global state.
Consequently,  correct  implementations may perform sequences of commands---and hence generate logs---that are
 different from those derived from the corresponding
protocol. In such cases, we still expect machines to be able to eventually agree on an
 interpretation of the log that matches one possible execution of the specification.

We tackle this problem by first defining the relevant events of an execution, namely those that are part of the \emph{effective log}.
Based on this, we establish an equivalence relation on logs that allows us to characterise the logs that can be produced by an execution of a swarm protocol's realisation as a swarm.
Armed with these tools we then state that all correct realisations produce valid effective logs and that all swarm protocol executions have corresponding swarms that realise them.

\subsection{Eventual fidelity}
We start by introducing some machinery for making precise the notion of correct implementation of a swarm protocol.

Roughly, one may think that $(\asysrt,\emptylog)$ is a faithful
implementation of a swarm protocol $\agt$ if it produces only global
logs that can be generated by $\agt$.
However, this notion is too strong for our setting; in fact, we appeal to a
weaker notion of fidelity such that for any global log $\alog$
produced by $(\asysrt,\emptylog)$,
i.e. $(\asysrt,\emptylog) \weakred (\asysrt,\alog)$,
there is a {\em related} log $\alog'$ that $\agt$ admits, i.e.
$(\agt,\emptylog) \weakred (\agt,\alog')$.
We postpone for a moment the formal definition of the expected
relation between logs, and convey some intuitions in the following
example.

\begin{example}\label{ex:intuition-efftypes}
Consider the swarm protocol $\agt$ in \cref{ex:taxi-service}, and the swarm $\mksys{P[],T[],T[],O[],T[]}\emptylog$ having three taxis dubbed $A$, $B$, and $C$. The swarm can produce\footnote{The details on how to reach such global log are in \cref{sec:full-example}.} the global log
	 \[
		\alog[\text{auc}]=\mklog{\aevent[][Requested],\aevent[B][Bid], \aevent[B][BidderID], \aevent[A][Bid], \aevent[A][BidderID], \aevent[][Selected], \aevent[C][Bid], \aevent[C][BidderID], \aevent[][passengerID]}
	 \]
	 Contrastingly, $\agt$ cannot generate such log; in fact, the protocol continuation after generating the prefix $\alog[1] = \mklog{\aevent[][Requested], \aevent[B][Bid], \aevent[B][BidderID], \aevent[A][Bid],\aevent[A][BidderID]}$ is $\Stchange(\agt, {\alog[\text{auc}]}) = \agt[\text{Bid}]$; hence, log $\alog[1]$ can only grow by appending $\mklog{\aevent[C][Bid],\aevent[C][BidderID]}$ or $\mklog{\aevent[][Selected], \aevent[][PassengerID]}$.
In the second case, we obtain the log  $\alog[2] = \mklog{\aevent[][Requested], \aevent[B][Bid], \aevent[B][Bidder], \aevent[A][Bid], \aevent[A][Bidder], \aevent[][Selected], \aevent[][PassengerID]}$.
Remarkably, all the machines discard the events $\aevent[C][Bid]$ and  $\aevent[C][BidderID]$ when processing  $\alog[\text{auc}]$, i.e., they
behave as if they were processing $\alog[2]$. In fact,  $\stchange(\afish[][P],\alog[\text{auc}]) = \stchange(\afish[][P],\alog[2])$,
$\stchange(\afish[][T],\alog[\text{auc}]) = \stchange(\afish[][T],\alog[2])$ and
$\stchange(\afish[][O],\alog[\text{auc}]) = \stchange(\afish[][O],\alog[2])$.
\finex
\end{example}
As highlighted by the previous example, despite the actual log generated by the swarm differing from  the logs generated by the protocol, all the machines are able to consistently discard those ill-generated events after complete propagation. In other words, the states of the machines in the swarm  depend only on a subset of the events in the log.
We characterise such subset via a type, called {\em effective type}. Intuitively, the effective type of a log is the type of the sublog containing all those events that are effectively relevant to the machines in the swarm.
Given a protocol $\agt$ and a subscription $\reactsto$, we expect an implementation to process only those events which some role has been subscribed to; consequently, our notion of \emph{effective type} is relative to a subscription.
However, the effective type of a log is not just the projection of its type with respect to the image of the subscription $\reactsto$. This is illustrated in the following example.
\begin{example}[Effective type and projection]
  The type of the log $\alog[\text{auc}]$ in \cref{ex:intuition-efftypes} is
  \[
	 \alogtype[\text{auc}] = \mklogtype{Requested, Bid, Bid, Selected, Bid, PassengerID}
  \]
  which differs from the type of $\alog[2]$ which is $\mklogtype{Requested, Bid, Bid, Selected, PassengerID}$.
Since the passenger $\arole[][P]$ processes all the types in $\alogtype[\text{auc}]$, we expect an appropriate subscription $\reactsto$ to be defined such that $\reactsto(\arole[][P]) \supseteq \{\aeventtype[][Requested], \aeventtype[][Bid], \aeventtype[][Selected], \aeventtype[][PassengerID]\}$.
Consequently, if we just keep the sublog of $\alogtype[\text{auc}]$ containing all types for which at least one role has been subscribed to, then we obtain exactly $\alogtype[\text{auc}]$, which does not reflect the type of the sublog that is effectively processed by the machines.
For this reason, the effective type depends also on the protocol being implemented.
\finex
\end{example}

\begin{definition}[Effective type]\label{def:eff}
    Let $\agt$ be a swarm protocol and $\reactsto$ a subscription.
The \emph{effective type of a log $\alog$ with respect to $\agt$
      and $\reactsto$},
written $\efftype=\efftype[@][@][@][\emptylog]$, is defined as follows
	\begin{align}
	  \efftype[\emptylog][@][@][\alogtype] &= \emptylog  \label{eq:effdone}
	  \\
	  \label{eq:effnonempty}
    \efftype[\aevent\logcat\alog][@][@][\emptylog] &= \aeventtype \logcat\efftype[\alog][\agt'][@][\alogtype']
    && \text{if}\quad
    \begin{minipage}[t]{.5\textwidth}
      \raggedright
      $\eventtyping$, $\aeventtype \in \reactsto(\roles(\agt,\reactsto))$, $\agt\red[{\gsumprefix[][@][{\aeventtype\logcat\alogtype}]}] \agt'$,
      and $\alogtype' = \filter[{\alogtype,\reactsto[{\activer[\agt']}]}]$
    \end{minipage}
    \\
    \label{eq:effpartial}
    \efftype[\aevent\logcat\alog][@][@][\aeventtype\gtpref\alogtype] &= \aeventtype\logcat\efftype[\alog][@][@][\alogtype]
    && \text{if}\quad\text{$\eventtyping$}
    \\
    \label{eq:effignore}
    \efftype[\aevent\logcat\alog][@][@][\alogtype] &= \efftype[\alog][@][@][\alogtype] && \text{otherwise}
  \end{align}
\end{definition}

As expected, the effective type of an empty log is the empty log type.
The effective type of  a non-empty log keeps track of  those
events that match the type of a log that can be generated by the protocol
(and discards all ill-ordered events). Hence,
the effective type of  $\aevent\logcat \alog$ with
respect to $\agt$ records the type $\aeventtype$ of the first event
$\aevent$ only if it has a  type that is expected by the protocol
(namely at least one of the roles in the protocol is subscribed to that type).
According to case~\eqref{eq:effnonempty}, a protocol
$\agt$ expects some event whose type $\aeventtype$ coincides with the guard of
one of its branches and  at least one
role is subscribed to $\aeventtype$. In such case, the effective type of the remaining log $\alog$
is processed first by consuming events of type $\alogtype'$ (rule~\eqref{eq:effpartial}), which
is the sequence of the remaining types generated by the branch that
are observed by the {\em active} roles in the continuation $\agt'$,
followed by considering the continuation $\agt'$.
We  remark here that only  the types of  events that are relevant for active roles
are reflected in the effective type (more details are given in~\cref{subsec:eff-proj}).
 Note that the
types that are not observed are just disregarded from the effective
type as per~\eqref{eq:effignore}.
We have that $\efftype$ is a well-defined function over deterministic
swarm protocols because log-determinism (\cref{def:gt-det}) ensures
that at most one branch of $\agt$ can match the event type
$\aeventtype$ of the event $\aevent$.

\begin{example}
  Let $\agt[\text{auction}]$ and $\agt[\text{choose}]$ the swarm protocol defined in
\cref{ex:taxi-service} and $\agt[\text{bid}] =
\gsumprefix[][Offer][Bid,BidderID][T]\gtpref\agt[\text{auction}]$.
We have
\begin{align}
  & \agt[\text{bid}] \red[{\gsumprefix[][Offer][{Bid,BidderID}][T]}] \agt[\text{Auction}]
  \label{eq:bid}
  \\
  & \agt[\text{auction}] \red[{\gsumprefix[][Select][{Selected,PassengerID}][T]}] \agt[\text{choose}]
  \label{eq:sel}
\end{align}
Let us compute the effective type of the log
$\alog = \mklog{\aevent[A][Bid], \aevent[A][BidderID],
  \aevent[][Rating], \alog'}$ on $\agt[\text{bid}]$ using a
subscription $\reactsto$ for which $\arole[][P]$, $\arole[][T]$ and $\arole[][O]$ are
subscribed to all events but $\aevent[][Rating]$.
We have
\begin{align*}
  \efftype[\alog][{\agt[\text{bid}]}] =\
  & \aeventtype[][Bid]\logcat\efftype[\mklog{\aevent[A][BidderID], \aevent[][Rating], \alog'}][{\agt[\text{auction}]}][@][{\aeventtype[][BidderID]}]
  \\
  = \ & \aeventtype[][Bid]\logcat\aeventtype[][BidderID] \logcat \efftype[\mklog{\aevent[][Rating], \alog'}][{\agt[\text{auction}]}][@][\emptylog]
  \\
  = \ & \aeventtype[][Bid]\logcat\aeventtype[][BidderID] \logcat \efftype[\alog'][{\agt[\text{auction}]}][@][\emptylog]
\end{align*}
where the first equality holds by \eqref{eq:effnonempty} since
$\aeventtype[][Bid] \in \reactsto(\arole[][T])$ and
\eqref{eq:bid}, the second equality holds by \eqref{eq:effpartial} since
$\eventtyping[][{\aevent[A][BidderID]}][BidderID]$, and the third
equation holds by \eqref{eq:effignore} since by hypothesis
$\aevent[][Rating] \not\in \reactsto(\arole[][P]) \cup
\reactsto(\arole[][T]) \cup
\reactsto(\arole[][O])$.
If $\alog' = \emptylog$ then
$\efftype[\alog][{\agt[\text{bid}]}] = \aeventtype[][Bid]\logcat\aeventtype[][BidderID]$
by \eqref{eq:effdone}.

Suppose instead that
$\alog' = \mklog{\aevent[][Selected],\aevent[][PassengerID],
  \aevent[B][Bid], \aevent[B][BidderID]}$.
Then, similarly to the first two equations above (using \eqref{eq:sel}), we have
\[
  \efftype[\alog'][{\agt[\text{auction}]}][@][\emptylog] = \aeventtype[][Selected]\logcat\aeventtype[][PassengerID] \logcat \efftype[\mklog{\aevent[B][Bid], \aevent[B][BidderID]}][{\agt[\text{choose}]}][@][\emptylog]
\]
And, by \eqref{eq:effpartial}, $\efftype[\mklog{\aevent[B][Bid], \aevent[B][BidderID]}][{\agt[\text{choose}]}][@][\emptylog]
= \efftype[\emptylog][{\agt[\text{choose}]}][@][\emptylog] = \emptylog$.
\finex
\end{example}

The relation between logs of an implementation with those of a
specification that we need is the equivalence induced by the equality
of their effective types.
\begin{definition}[Log equivalence]
  Two logs $\alog$ and $\alog'$ are \emph{equivalent} with respect to
 a swarm protocol  $\agt$ and a subscription  $\reactsto$,
  written $\eqlog\alog{\alog'}$, if they have the same
  effective type with respect to $\agt$ and $\reactsto$, i.e., $\efftype = \efftype[\alog']$.
\end{definition}

Then, the notion of correct implementation is simply stated as follows.

\begin{definition}[Eventual fidelity]
A swarm $(\asysrt,\emptylog)$ is {\em eventually faithful}  to a  swarm protocol $\agt$ and
a subscription $\reactsto$ if $(\asysrt,\emptylog) \weakred (\asysrt,\alog)$ implies there exists $\alog'$ such that
 $(\agt,\emptylog) \weakred (\agt,\alog')$ and $\eqlog\alog{\alog'}$.
\end{definition}

\subsection{Implementation correctness by projection}
Our projection operation yields an effective procedure for obtaining
correct implementations out of well-formed swarm protocols which we call
\emph{realisations}.

\begin{definition}[Realisation]\label{def:realisation}
Let $\agt$ be a swarm protocol and $\reactsto$ be a subscription.
A \emph{realisation (of size $n$) of $\agt$ with respect to
	 $\reactsto$}, shortened as \emph{\realise}, is a swarm
  $(\asysrt,\emptylog)$ of size $n$ such that, for each
  $1 \leq i \leq n$, there exists a role $\arole[]\in\roles[\aspecdef]$
  such that $\asysrt(i) = (\agt\aproj[{\arole[]}], \emptylog)$.
A realisation $\asysrt$ is \emph{complete} if for all
  $\arole[]\in\roles[\aspecdef]$ there exists $1 \leq i \leq n$ such that
  $\asysrt(i) = (\agt\aproj[{\arole[]}], \emptylog)$; we call
  \emph{partial} a realisation which is not complete.
\end{definition}
Remarkably, the number of machines in a realisation is not related to
the number of roles in the corresponding swarm protocol.
Indeed, \cref{def:realisation} simply requires that each machine in
the swarm plays one of the roles in the swarm protocol.
Concretely, we may have several components implementing the same role
(i.e., the role is replicated) as well as roles without a
corresponding machine, that is partial realisations.

\begin{example}[Realisations]
  The swarm protocol for the running example defined in
  \cref{ex:taxi-service} would typically be realised by one machine for the
  passenger $\agt\aproj[{\arole[][P]}]$, several taxis running
  the machine $\agt\aproj[{\arole[][T]}]$, and at least one accounting office running $\agt\aproj[{\arole[][O]}]$.
A partial realisation could be one without an accounting office, in which case no machine can generate $\aeventtype[][receipt]$ events.
\finex
\end{example}

The remainder of this section is devoted to showing that realisations (either complete or partial)
are eventually faithful if they are obtained by projecting well-formed swarm protocols.

\subsubsection{Projections and effective types}\label{subsec:eff-proj}
We  first establish a correspondence between the behaviour of a single projection and that of
the respective protocol. In particular,  we show that
effective types provide an accurate abstraction of the
information   contained in a log that is relevant for a role. Concretely, the next result states that a projection enables  a command
after processing a log $\alog$ only when the protocol enables the
same command after producing an equivalent log $\alog'$.
\begin{restatable}{lemma}{restateCmdEnabledAdmissibleLog}\label{lem:cmd-on-admissible-log}
If $\agt$ is a \wf\ swarm protocol and
    $\obscmd[\stchange(\agt\aproj,\alog)]$
then there exists $\eqlog{\alog'}\alog$ such that
  $(\agt,\emptylog) \weakred[] (\agt,\alog')$ and
  $\Stchange(\agt,\alog') \red[{\gsumprefix[][@][{\alogtype}]}] \agt'
  $.
\end{restatable}
One might think that equivalent logs are indistinguishable for a machine, i.e.,   that
  $\eqlog\alog{\alog'}$ implies $\stchange(\agt\aproj,\alog) = \stchange(\agt\aproj,\alog')$.
  However this might not be the case if logs are incomplete, in the sense that they do not include
  all the events  generated by the same command.

\begin{example}[Log equivalence does not imply indistinguishable states]\label{ex:swithindependent}
Consider the swarm protocol $\agt = \gsumprefix[][c][a,b][R]\gtpref \gsumprefix[][d][c][S]\gtpref \zero$ with $\reactsto=\bigl\{    \arole[][R]\mapsto\{\aeventtype[][a],\aeventtype[][b]\},    \arole[][S]\mapsto\{\aeventtype[][a],\aeventtype[][c]\}\}$.
If $\eventtyping[][{\aevent[][a]}][{\aeventtype[][a]}]$ and
$\eventtyping[][{\aevent[][b]}][{\aeventtype[][b]}]$ then $\efftype[{\aevent[][a]}] = \aeventtype[][a]$ and $\efftype[{\aevent[][a]\logcat\aevent[][b]}] = \aeventtype[][a]$;
in fact the first equation holds by definition and the second holds because  $\aeventtype[][b]\not\in\reactsto[{\activer[{\gsumprefix[][d][c][S]\gtpref \zero'}]}] = \reactsto[{\arole[][S]}] = \{\aeventtype[][a],\aeventtype[][c]\}$.
Therefore, $\eqlog{\aevent[][a]} {\aevent[][a]\logcat\aevent[][b]}$.
Now take the projection of $\agt$ over  $\arole[][R]$  with respect to $\reactsto$, i.e., $\afish[{\arole[][R]}] = \agt\aproj[{\arole[][R]}] = \inp[][a] \inp[][b] (\gsumprefix[][d][c][S]\gtpref \zero\aproj[{\arole[][R]}]) =  \inp[][a] \inp[][b] \zero$.
Clearly $\stchange(\afish[{\arole[][R]}], \aevent[][a]\logcat\aevent[][b]) \neq  \stchange(\afish[{\arole[][R]}] ,\aevent[][a])$.
\finex
\end{example}

Two considerations on \cref{ex:swithindependent} are worthwhile.
On the one hand, while $\aevent[][a]\logcat\aevent[][b]$ has all the events produced by the
execution of the command $\acmd$, the log consisting of just the event $\aevent[][a]$ does not. Since we assume that all events are eventually propagated, our technical development in the next section will disregard incomplete (global) logs.
On the other hand, one may wonder about the fact that effective types do not collect information of events that are not observed by active roles. This is essential to account for the fact that a realisation may interject events.
For instance, a realisation may actually generate a log of type $\mklog{\aeventtype[][a], \aeventtype[][c], \aeventtype[][b]}$ because a machine that implements the role $\arole[][S]$ may perform the command $\acmd[][d]$ as soon as it processes an event of type $\aeventtype[][a]$; hence the generated  event can precede the one of type $\aeventtype[][b]$ in the consolidated log.
If our notion of log equivalence were fine enough to distinguish logs of type $\mklogtype{a, c, b}$ from $\mklogtype{a, b, c}$ then we would rule out
implementations behaving as above, which is not what we want because the interaction does not violate the protocol.

\subsubsection{Characterisation of the logs admitted by a protocol}

We now provide a characterisation of the logs that can be generated by
a realisation of a well-formed swarm protocol.
To do this we have to take into account for the possible reordering
and the spurious events that can be generated by machines that
faithfully implement a protocol.
Intuitively, we may think that a realisation generates logs that correspond to the
combination of several executions of the protocol, which might  share a common prefix.
Consider again the swarm protocol $\agt$ in \cref{ex:taxi-service}. As discussed in \cref{ex:intuition-efftypes}, we expect realisations to be able to generate logs such as $\alog[\text{auc}]$ of this example.
Note that such log can be generated by merging, among others, two different reductions of $\agt$, e.g. $(\agt,\emptylog) \weakred[] (\agt,\mklog{\aevent[][Requested], \aevent[B][Bid], \aevent[B][BidderID], \aevent[A][Bid], \aevent[A][Bidder], \aevent[][Selected], \aevent[][PassengerID]})$ and $(\agt,\emptylog) \weakred[]
(\agt,\mklog{\aevent[][Requested], \aevent[B][Bid], \aevent[B][Bidder], \aevent[C][Bid], \aevent[C][BidderID]})$.
Note that the reductions share events (accounting for an scenario in which the computation has diverged).
Intuitively, two runs can be combined either if  they produce disjoint logs
or they share events that  come from a common execution prefix (as in the previous example). Formally, two runs $(\agt,\emptylog) \weakred[] (\agt,\alog[i])$ with $i  =1, 2$ are {\em consistent} if there exist logs $\alog$, $\alog[1]'\cap\alog[2]' = \emptyset$, such that $\alog[i] = \alog\logcat\alog[i]'$ and $(\agt,\emptylog) \weakred[] (\agt,\alog) \weakred[] (\agt,\alog\logcat\alog[i]')$ for $i = 1,2$.
The notion of consistency is lifted to sets of runs $\{(\agt,\emptylog) \weakred[] (\agt,\alog[i])\}_{1 \leq i \leq k}$, by requiring pair-wise consistency.
We write $\alog[i]^j$ for the sequence of events produced by the
$j$-th step in the reduction $i$, i.e.,
$(\agt,\emptylog) \weakred[]^{j-1} (\agt,\alog[i]')\red[{\afishact}](\agt,\alog[i]'\logcat\alog[i]^j)\weakred[] (\agt,\alog[i])$.

\begin{definition}[Admissible log]
  A log $\alog$ is \emph{admissible} for a \wf\ protocol $\agt$ if
  there is a set of consistent runs
  $\{(\agt,\emptylog) \weakred[] (\agt,\alog[i])\}_{1 \leq i \leq k}$
  and a log $\alog' \in (\mergelog_{1 \leq i \leq k} \alog[i])$ such
  that $\eqlog\alog {\alog'}$,
  $\alog = \bigcup_{1 \leq i \leq k} \alog[i]$, and for all
  $1 \leq i \leq k$ and $\alog[i]^j\sublog\alog[]$ for all events
  $\alog[i]^j$ produced by the $j$-th step in the reduction $i$.
\end{definition}

Remarkably, admissible logs are not just those that can be
obtained by merging several logs $\alog[i]$; it may be the
case that $\alog$ is admissible but
$\alog \not\in (\mergelog_{1 \leq i \leq k} \alog[i])$.
In fact the notion is weaker and accounts for the possible reorderings of events
that do not change the effective type of the log.
Consider the protocol  in \cref{ex:swithindependent} and its complete realisation consisting of two machines.
As previously discussed, that realisation may generate a log of type $\aeventtype[][a]\logcat\aeventtype[][c]\logcat\aeventtype[][b]$.
With a single run of the protocol, i.e.~by fixing $k=1$ and taking
$(\agt,\emptylog) \red[{\afishact[][@][{\aeventtype[][a]\logcat\aeventtype[][b]}]}] (\agt,\aevent[][a]\logcat\aevent[][b])
\red[{\afishact[][d][{\aeventtype[][c]}]}] (\agt,{\alog[1]})$ with $\alog[1] = \aevent[][a]\logcat\aevent[][b]\logcat\aevent[][c]$,
 we can conclude that $\aevent[][a]\logcat\aevent[][c]\logcat\aevent[][b]$ is
generated by some realisation. Note that $\mergelog_{1 \leq i \leq 1} \alog[i] = \{\alog[1]\} = \{\aevent[][a]\logcat\aevent[][b]\logcat\aevent[][c]\}$.
Hence, $\alog' \in \mergelog_{i \leq 1 \leq 1} \alog[i] $ iff
$\alog' =  \aevent[][a]\logcat\aevent[][b]\logcat\aevent[][c]$ and $\efftype[{\alog'}] = \aeventtype[][a]\logcat\aeventtype[][c]$.
Then, the log $\aevent[][a]\logcat\aevent[][c]\logcat\aevent[][b]$ is equivalent (i.e., it has the same effective type),
 and moreover it has the same elements and preserves the relative order between events generated by the same command  (i.e., $\aevent[][a]$ precedes $\aevent[][b]$).
 Hence, we conclude that the protocol admits  the log $\aevent[][a]\logcat\aevent[][c]\logcat\aevent[][b]$.
On the contrary,  the last condition $\alog[i]^j\sublog\alog[]$ about the preservation of the relative order of events generated by the
same command bans logs such as $\aevent[][b]\logcat\aevent[][a]\logcat\aevent[][c]$.

The following result ensures that any admissible log of a well-formed protocol is equivalent to a log obtained by the sequential execution of the protocol.
In other words,  despite a log may contain events produced by decisions that are in conflict, its effective type corresponds to a sequential run.

\begin{restatable}{lemma}{restateAdmissibleReduction}\label{lem:admissible-reduction}
  If $\alog$ is admissible for a \wf\ swarm protocol $\agt$ then there
  exists a log $\alog'$ such that
  $(\agt,\emptylog) \weakred[] (\agt,\alog')$ and
  $\eqlog{\alog}{\alog'}$.
\end{restatable}

Moreover,  if we extend an admissible log with events generated by the execution of command enabled over a partial view of the global log, then
we obtain an admissible log. This property is instrumental for our main result in the following section
(\cref{th:main}).

\begin{restatable}{lemma}{restateAdmissibleMerge}\label{merge-admissible}
  Let $\alog[1]$ and $\alog[2] \subseteq \alog[1]$ be admissible logs
  for a \wf\ swarm protocol $\agt$.
If
  $(\agt,\alog[2]) \red[{\gsumprefix[][@][{\alogtype}]}] (\agt,\alog[2] \logcat \alog[3])$
  and $\alog\in\alog[1]\mergelog(\alog[2]\logcat \alog[3])$ then
  $\alog$ is admissible for $\agt$.
\end{restatable}

\subsubsection{Realisations are faithful}
We now address our main result, which shows that  a realisation
of a well-formed protocol only generates  admissible logs.

\begin{restatable}{theorem}{restateSerializability}\label{th:main}
  Let $(\asysrt,\emptylog)$ be a realisation of a \wf\ swarm protocol
  $\agt$.
If $(\asysrt,\emptylog) \weakred (\asysrt',\alog)$ then $\alog$ is
  admissible for $\agt$.
\end{restatable}

Since every admissible log is equivalent to a log generated by
the  protocol  (\cref{lem:admissible-reduction}), we conclude that any realisation of a well-formed swarm protocol is  eventually
faithful (i.e., correct).
Note that this implies that all realisations of a well-formed swarm protocol exhibit eventual consensus~\cite{8919675} regarding which branch is taken in its choices (concretely, there is a $\red[\tau]$ step after which all machines take the same branch in their state computation).

\begin{restatable}{corollary}{restateFidelitycorollary}\label{cor:faithful}
  Every  realisation of size $n$ of a \wf\
  swarm protocol $\agt$ is eventually faithful with respect to $\agt$ and $\reactsto$.
\end{restatable}

The above result is independent from the number of replicas that implement each role; it also
holds for partial realisations (i.e., when some roles are absent).

\subsection{Implementation completeness}
Differently from common  session type systems, the behaviour of a complete realisation (i.e., one in which every role
is implemented) is complete with respect
to the  protocol, in the sense that every reduction of the protocol can be mimicked by the realisation.
This derives from the fact that  non-determinism  in our model  arises from the
execution of external commands but not because of the abstraction of internal (and customary deterministic) choices.
Firstly, we note that logs that are generated sequentially according to the protocol drives the
machines to the corresponding states.
\begin{restatable}{lemma}{restateProjMimick}\label{lem:projpres}
If $\agt$ is \wf\ swarm protocol and $(\agt,\emptylog) \weakred[] (\agt,\alog)$ then
  $\stchange(\agt\aproj,\alog) = \stchange(\agt,\alog)\aproj$ for all
  $\arole[]\in\roles[\aspecdef]$.
\end{restatable}
Moreover, \cref{lem:complete} below states that a complete realisation
is able to generate the logs that are generated by the protocol (the result is obtained
by using previous result and
by propagating all events to all replicas right after a machine
performs a command).
\begin{restatable}{proposition}{restateRealisationCovers}\label{lem:complete}
Let $(\asysrt,\emptylog)$ be a complete realisation of size $n$ of the \wf\ swarm protocol $\agt$.
   If $(\agt,\emptylog)\weakred (\agt,\alog)$ then
  there is a swarm $\asysrt'$ such that $(\asysrt,\emptylog) \weakred (\asysrt',\alog)$.
\end{restatable}

\subsection{Deadlocks, liveness, and safety}\label{sec:deadlockfree}

Given our co-inductive definition of swarm protocols $\agt=\agsum$ it is obvious that every protocol state $\agt'$ that is a subterm of $\agt$ can be reached by a suitable sequence of command invocations.
A complete realisation---which after sufficient event propagation allows us to invoke any enabled command---can therefore reach local machine states corresponding to $\agt'$ as per \cref{lem:complete}.
In general, \emph{liveness} is not guaranteed in our theory because it permits state cycles to be visited without bound.
We conjecture that the usual approach of imposing a suitable notion of fairness could guarantee liveness.
Finding such conditions for our system is still an open problem.

Assuming eventual event propagation (i.e. $\red[\tau]$ will eventually be exhausted) and assuming that every machine with enabled commands will eventually choose to invoke one of them, it is also obvious that progress will always be made as long as the swarm protocol permits it (i.e. as long as the swarm protocol state induced by the effective global log enables commands).
In other words, our swarms are \emph{deadlock-free}.

We take a different approach from session typing disciplines in that orphan messages are deliberately allowed---our event processing semantics ensure that unexpected messages are dropped by the receiver.
Together with our well-formedness conditions we therefore ensure that each machine only processes exactly those messages that it needs to faithfully progress according to the global protocol, as detailed by its partial view as long as events are still propagating.

\section{Related work}\label{sec:rw}
It is widely accepted that solutions to distributed coordination
problems strongly depend on the adopted computation
model~\cite{gay2017behavioural,hlvccdmprt16,ancona2016behavioral}.
Our proposal is grounded on the principles of \emph{local-first
  cooperation}~\cite{lfc,localfirst}.
Key to this architecture is the \emph{autonomy} of each
participating node within a swarm.
Autonomy allows each node to make progress independently of network
connections, availability of other nodes, or delay in the
communications.
Our target model features specific hues that distinguish it from
other behavioural types systems.
In our case, distributed heterogeneous components interact
asynchronously by emitting and consuming \emph{events}
according to a role specified in a given protocol (such as passenger and the taxis
in our running example).
More precisely, events are the side effects of commands
non-deterministically executed by components; events are locally
logged by each of the components and asynchronously spread through the
swarm.
Crucially, we do not make any assumption on the relative speed of
communications and simply require that logs  \emph{eventually} agree
on the order of events~\cite{burckhardt2014principles}.
This liberal setting permits inconsistencies: components may take
discordant decisions which compromise the execution of the protocol
and exhibit behaviour precluded in strongly consistent models.
Our approach lies within methods related to data replication, which
are notoriously complex.
In fact, standard techniques have to trade-off among availability,
consistency, and partition tolerance~\cite{CAP}.
Several techniques have been proposed, such as conflict-free replicated
data types~\cite{DBLP:conf/sss/ShapiroPBZ11}, cloud
types~\cite{burckhardt2012cloud}, consistency
contracts~\cite{sivaramakrishnan2015declarative},
invariants~\cite{gotsman2016cause,kaki2018safe,DBLP:journals/pvldb/BalegasDFRP18}, linearizability~\cite{wang2019replication}, and
operational models for applications such as
GSP~\cite{DBLP:conf/ecoop/BurckhardtLPF15,gotsman2017consistency}.
An original facet of our approach is that we use behavioural types to
discipline data replication in order to eventually reach consistency.
We focus on the consequences that arise from the ability of each node
to take decisions based exclusively on \emph{local information}.

Our proposal is inspired by the choreographic framework introduced in
the seminal work on \emph{multiparty session types}
\cite{honda2008multiparty,honda2016multiparty}.
However, the peculiarities of our execution model as well as on the
properties that we target require a radical change in the definition
of well-established notions of
global types, such as projections and well-formedness.
The main originality of our approach is that components speculatively
proceed along several (possibly inconsistent) branches of distributed
choices provided that an agreement is eventually reached.
Intuitively, this is attained by disregarding all executions bar
one when the local logs \quo{consolidate}, namely when relevant
events have propagated to all relevant components.
As far as we are aware of, multiple selectors are forbidden in the
well-formedness conditions of most behavioural type
systems~\cite{hlvccdmprt16}.
A slight weakening of this condition is given in~\cite{jy20} but the
conditions there still reject the protocol in~\cref{ex:local-bidding}.
Noteworthy, we divert from the research path of
behavioral types with respect to the properties we are after.
We aim to guarantee that projections of global specifications yield
realisations of swarms that eventually reach a consistent view of the
distributed execution, even in presence of transitory inconsistencies.
This is in contrast with behavioural type systems designed to attain
(dead)lock freedom or some notions of progress
(see~\cite{hlvccdmprt16} and~\cite{dp22} for a recent account on the
binary case).

Secondly, our behavioural specifications completely abstract over the number of instances that execute each role of a protocol.
This is often not the case for multiparty session types. 
Parametric multiparty session types have been considered in
\cite{ydbh10,dy11,mybh12} and more recently in~\cite{chsny19,jy20}.
These proposals aim to capture the fact that roles in a protocol are
\quo{connected} to form a topology that can be generalised (e.g.
parameterising a ring topology by its size).
These behavioural type systems therefore require to explicitly handle
the parameters of the protocol.
Our specifications are instead completely oblivious of such parameters.
To the best of our knowledge, multiparty session types have focused on
point-to-point, message-passing communication model, even to deal with
highly dynamic scenarios, as those involving robot
coordination~\cite{majumdar2019motion}.

\section{Final Remarks}\label{sec:conc}

The behavioural types proposed here are rather unconventional.
We deviate from the default choice of implementing communication
protocols on top of point-to-point, message-passing communication,
which is a common practice (see e.g.~\cite{dd09,vas12,hlvccdmprt16}).
Components in our setting interact via a shared log that is
distributedly built without any further coordination mechanism.
More precisely, each component keeps a local, possibly partial and
inconsistent view of the global log.
Based on that view alone a component
may perform an action with immediate effect on its local state;
those effects are then propagated asynchronously to the rest of the
system.
This implies that components can perform
globally invalid actions (as long as they are locally valid),
but we require every component involved in the protocol to be able
to recognise these and eventually behave as if only valid actions were
performed.
Technically this means that we renounce established properties like \emph{session fidelity},
instead focusing on the ability of the system to eventually agree without dedicated coordination;
we do note that our typing discipline guarantees deadlock-freedom, though.

Our target applications intrinsically involve sets of components whose number is
 statically unknown: components may dynamically join and leave the
execution of an interaction.
As is common practice in behavioural types, we describe protocols in terms of the roles that components
may play.
Unlike most behavioural types, ours are agnostic to the number of
instances playing each role.
We assume that any role in any swarm protocol can be replicated as many
times as needed.
This choice also impacts on the interpretation of choices.
In standard approaches, every choice is assigned to a single role
implemented by a single component responsible for coordinating the
decision.
This is problematic when the implementation of a role can be
replicated, even more when the states of the replicas may be
misaligned: different components may decide differently.
Consequently, choices in our swarm protocols are intended to be resolved
distributedly among components that may implement different roles.
Our solution is based on speculative computation: different choices
can proceed concurrently until components are able to agree (by inspecting their local state) on the branch that has been
selected based on a total order for all events (implemented for example using Lamport timestamps and unique node identifiers).

The computational model presented in this paper gives an abstract
description of an existing infrastructure~\cite{actyx}, in which machines are
actually implemented as programs in a mainstream programming language
(e.g.~{\tt TypeScript}).
In this respect, the machines presented here play the role of
local types that describe the intended behaviour of each component.

Determining a suitable subscription could be hard in general.
We conjecture that the swarm protocol itself could be used to infer a
suitable \quo{minimal} registration enforcing well-formedness to be
suggested to designers.
Such subscription could then manually be refined, provided that well-formedness
is preserved.
Moreover, we may envision programmers specifying just the relevant
information that needs to be transmitted and then automatically infer
the events needed for coordinating choices (pretty much in the style
of the communication of labels in session types).

An underlying assumption about speculative execution is that the
effects of performing invalid actions can be discarded. In other
words, invalid actions have no consequences.
In several situations this may be unacceptable.
Swarm protocols could be used to systematically identify
such situations---e.g.\ by noting when corresponding events are disregarded by machines---and
enable principled treatment at the application level.
We plan to study suitable extensions for our
projections that automatically inject the required behaviour
for executing amending actions.
Alternatively, we will explore monitoring approaches equipped with
sanitisers responsible for compensation.

We have only partly addressed failures in our model: while we do model transient inability to receive (which would inhibit the event propagation transition \rulename{Prop}) or to operate (i.e.\ inhibit command invocation \rulename{Local}), we do not model permanent inability to send.
In the presence of a stop failure a machine could communicate the first event of a choice but then fail in propagating all the expected following events resulting from the command invocation, in which case the system could get stuck.
A fix for this issue could be to only proceed with an external choice once all specified events are present in the local log, allowing the swarm to permanently discard a choice made by a failing machine; this could be expressed by ingesting logs instead of single events in the definition of machine semantics.

Another extension that deserves further study has been hinted at in \cref{sec:machine-interaction}:
it would make sense and achieve a per-choice notion of non-interference if the first event of a choice not only decided which branch to take but also from which source machine to consume the rest of the choice's events.
This would further strengthen the failure handling sketched above by making sure that inputs from failing machines are consistently discarded.
Characterising the precise guarantees that derive from such a scheme will be an interesting topic for further study.

\pagebreak
\appendix
\section{Supplementary material}\label{sec:app}

\subsection{Machine syntax}\label{sec:formal-machine}

Throughout the paper we fix a set \(\eventset\) of \emph{events} (ranged
over by \(\aevent, \aevent[1], \aevent[2], \dots\)) and define a
\emph{log} $\alog$ as a sequence of events without repetitions; we
denote the set of logs as $\logset$.
We assume the existence of a decidable typing relation
\(\eventtyping\) that uniquely assigns a type \(\aeventtype\) to
\(\aevent\), with \(\typeset\) being the set of event types.
Consequently, logs can be typed by assigning a \emph{log type}
\(\alogtype \in \typeset^\star\), which is a sequence of event types.
We will let a sequence denote its underlying set of elements; for instance, we write $\aevent \in \alog$ when $\aevent$ occurs in $\alog$.
For a natural number $n$, we let $\natseg$ denote the set $\{1, \ldots, n\}$ (note that $\natseg[0] = \emptyset$).
In the following we only consider deterministic machines.

\begin{definition}[Determinism]\label{def:det}
  A machine
  $\afish = \asumnew[@][{\areaction[1], \cdots, \areaction[n]}]$ is
  \emph{deterministic} if event types are pairwise different between
  branches, i.e. $\aeventtype[j] \neq \aeventtype[k]$ for all
  $j \neq k \in \natseg$, and
  $\afish[i]$ is deterministic for all $i \in \natseg$.
  The class of deterministic machines is denoted by $\apond$.
\end{definition}

We will specify machines as finite sets of equations $\afish=\dots$ where the occurrences of $\afish$ on the right-hand side are prefix-guarded.
A regular tree satisfying such equations is guaranteed to be unique and corresponds to a finite automaton~\cite{Courcelle83}.
Our textual presentation is equivalent to the one based on automata used in the main part\footnote{
For the sake of our results, it is immaterial how to obtain a finite-state automaton from a regular tree; the interested reader is referred to \cref{sec:regtrees2lts} for the details.
}
that we will occasionally exploit since it is useful both for visualisation and for reasoning about decidability and complexity of the properties of machines.
We will also sometimes refer to $\afish$ as the \emph{state} of a finite-state automaton.

The behaviour of a machine is determined by its partial view of the execution, i.e. its local log.
Hence the operational semantics are based on log processing.
We let $\rdy(\afish)$ be the set of event types that machine $\afish$ handles in its initial state:
\begin{align*}
  \rdy[\afish] =
   \{\aeventtype[1], \ldots, \aeventtype[n]\} \qand[for] \afish = {\asumnew[@][{\areaction[1], \cdots, \areaction[n]}]}
\end{align*}

The current state \(\stchange(\afish,\alog)\) of a machine $\afish$
with respect to a log $\alog$ is given by a map $\stchange$ that returns
the machine $\afish'$ reached after consuming the events in $\alog$.
Formally,
\begin{align}\label{eq:delta1}
 \stchange(\afish,\emptylog) = &\ \afish \\
\label{eq:delta2}  \stchange(\afish,\aevent \cdot \alog) = &
    \begin{cases}
      \stchange(\afish, \alog)
      &
        \text{if }\  \eventtyping \text{ and } \aeventtype \not\in \rdy(\afish)
      \\
      \stchange(\afish[j], \alog)
      &
        \text{if }\  \eventtyping[][@][{\aeventtype[j]}]  \text{ and } \afish = \asumnew[@][{\areaction[1], \ldots,\areaction[j], \ldots, \areaction[n]}]
     \end{cases}
\end{align}
It is a simple observation that $\stchange$ is a well-defined function for deterministic machines.
A peculiarity of our state transition function is that it ignores events when they do not match any of the guards in the current state.

The capability of a machine $\afish$ to execute a command $\acmd$ that emits a non-empty sequence of events of type $\alogtype$ is written $\obscmd$; we say that $\acmd$ is enabled at $\afish$ if $\obscmd$ holds.
For a machine $\afish=\asumnew[@][\ldots]$ this is the case if
$\acmdrel(\acmd[])=\alogtype[]$.
Command invocation (which is an external stimulus in our model) proceeds
according to the following rule:
\begin{align}\label{eq:cmd}
  \mathrule{{\stchange(\afish,\alog)} = \afish' \qquad \obscmd[\afish'][@][\alogtype] \qquad
  \eventtyping[][\alog'][\alogtype]  \qquad \alog' \in \logset\ \text{ fresh}}{
    (\afish, \alog) \red[{\afishact}] (\afish, \alog\cdot\alog')
  }{Cmd}
\end{align}
One noteworthy detail is that the emitted events are fresh, i.e. distinct from any other events emitted prior, concurrently, or afterwards.
In practice this is achieved by including an origin timestamp, e.g. a logical or physical clock timestamp paired with a unique network node identifier.
Also note that the fresh events are appended to the local log, ordering them \emph{later than} all events currently known at this location.
The state of the machine after applying the \rulename{Cmd} rule is given by \(\stchange(\afish,\alog\cdot\alog')\), establishing that changes in the local log are processed before the next command can be enabled.
This allows control over how often a command can be invoked.

\begin{restatable}{lemma}{restateLocalRedLog}\label{prop:local-red-log-extension}
  If $(\afish,\alog) \red[{\afishact}] (\afish,\alog')$
  then there is $\alog'' \in \logset$ such that
  $\eventtyping[][\alog''][\alogtype]$ and
  $\alog' = \alog \logcat \alog''$.
\end{restatable}
\begin{proof}
  Trivially from rule \rulename{cmd} observing that the log is
  extended with fresh events.
\end{proof}

\subsection{Formalising swarms }\label{sec:formal-swarm}
We now introduce the model that governs the interaction among
machines and across network nodes.
As described above, a machine acts on its local
log by reading it or by extending it with fresh events.
Changes propagate asynchronously to the local logs of other machines.
Consequently, the log currently available to each machine may contain just a partial
view of the set of events generated in the system so far.

\begin{definition}[Swarms]\label{def:rtsys}
  A \emph{swarm (of size $n$)} is a pair
  $(\asysrt,\alog) \in (\apond \times \logset)^{\natseg} \times
  \logset$, where \(\asysrt(i) = (\afish[i],\alog[i])\) describes the
  $i$-th machine.
Each event is assigned a \emph{source machine} by the map
  \(\source: \eventset \to \nat\), meaning that an event \(\aevent\)
  has been the result of a command invocation at machine
  \(\asysrt(\source(\aevent))\).
\end{definition}
Note that the same machine \(\afish\in\apond\) may be replicated, that is it can occur as the machine of multiple participants of a swarm;
each participating machine has a unique identity and tracks a separate local log
(intuitively, the identity of a participant is its index in the (domain of the) swarm $\asysrt$).
It is convenient to let $(\afish[1],\alog[1]) \parop \ldots \parop (\afish[n],\alog[n]) \parop \alog$ denote the swarm $(\asysrt,\alog)$
such that $\asysrt(i) = (\afish[i],\alog[i])$ for all $i \in \natseg$.
We dub $\alog$ the \emph{global} log of $\asysrt$ and $\alog[i]$ the \emph{local} log of machine $\afish[i]$.

\begin{definition}[Coherence]\label{def:wf-runtime}
  A swarm
  $(\asysrt,\alog)=(\afish[1],\alog[1]) \parop \ldots \parop \
  (\afish[n],\alog[n]) \parop \alog$ is \emph{coherent} if
  $\alog[i] \sublog \alog$ for all $i\in\natseg$ and
  \(\alog = \bigcup_{ i \in \natseg} \alog[i]\).
\end{definition}

This definition establishes a consistency requirement among local and global views:
each local log is a sublog of the global one, which in turn contains no events beyond the local logs.
Hereafter, we assume swarms to be coherent.
The condition of downward-completeness allows more concise characterisation of local logs in real implementations
since a sublog is uniquely described by the set of each source's latest event in the context of a global log.
In other words, a sublog can be represented by a set of event identities (e.g. source and timestamp) whose cardinality is at most the size of the swarm;
this is used in the Actyx implementation to tag state snapshots in order to speed up subsequent state computations.
Downward-completeness will also be crucial in the causal consistency rule given in \cref{def:cc}.

Note that, by coherence, all elements in $\alog[i]$ are already in $\alog$.
Consequently, the merge operation extends the global log $\alog$ with the newly generated events,
which can be interleaved at any position after the occurrence of the last element of $\alog[i]$ in $\alog$.

\begin{example}[Event generation]\label{ex:nd-prop}
  Assume that $(\afish[1],\aevent[@][b])\red[{\afishact}] (\afish[1],\aevent[@][b]\mdot\aevent[@][d]\mdot\aevent[@][e])$ holds.
  Then the swarm $(\asysrt,\alog)=(\afish[1],\aevent[@][b]) \parop \ldots \parop \aevent[@][a]\mdot\aevent[@][b]\mdot\aevent[@][c]$
  (with \(\aevent[@][a]\) and \(\aevent[@][c]\) from some other source) can evolve as
  \begin{equation*}\label{eq:ex-rule-local}
    (\asysrt,\alog)
    \red[\afishact]
    (\afish[1],\aevent[@][b]\mdot\aevent[@][d]\mdot\aevent[@][e]) \parop \ldots \parop \alog'
    \quad\text{with}\quad\alog' \in
    \{ \aevent[@][a]\mdot\aevent[@][b]\mdot\aevent[@][c]\mdot\aevent[@][d]\mdot\aevent[@][e],
    \
    \aevent[@][a]\mdot\aevent[@][b]\mdot\aevent[@][d]\mdot\aevent[@][c]\mdot\aevent[@][e],
    \
    \aevent[@][a]\mdot\aevent[@][b]\mdot\aevent[@][d]\mdot\aevent[@][e]\mdot\aevent[@][c]
    \}
  \end{equation*}
  We use non-determinism to model the fact that there are no guarantees about the relative order of the generated events with respect to those events that are locally unknown.
  \finex
\end{example}

Rule \rulename{Prop} establishes our policy for the propagation of
events.  A local log can be extended with events in the global log
$\alog$ provided that the extension is still a sublog of $\alog$; this
constraint ensures in-order delivery of events stemming from the same
command invocation.

\begin{example}[Event propagation]\label{ex:nd-prop2}
  Consider the coherent swarm
  \begin{equation*}
	  (\afish[0],\aevent[0][a]) \parop
    (\afish[1],\mklog{\aevent[1][b], \aevent[1][d], \aevent[1][e]}) \parop
    (\afish[2],\aevent[2][c]) \parop
    \mklog{\aevent[0][a], \aevent[1][b], \aevent[1][d], \aevent[2][c], \aevent[1][e]}
  \end{equation*}
  where subscripts represent the source of each event.  By rule
  \rulename{Prop}, we can propagate \(\aevent[@][a]\) to machine 2,
  yielding \(\aevent[@][a]\mdot\aevent[@][c]\)---note that this log may mean that the command that led to
  \(\aevent[@][c]\) should not have been enabled in the first place!
  We could also propagate \(\aevent[@][b]\) or the whole global log
  to machine 2, but we cannot propagate \(\aevent[@][d]\) without
  also propagating \(\aevent[@][b]\) as per \cref{ex:sublogs}.
  \finex
\end{example}

\subsection{A full-fledged example}\label{sec:full-example}

We illustrate the salient points of our framework on a complex scenario of our running example.
Consider the following set of machines for our running example:
\begin{align*}
  \text{passenger:}\quad\afish[][P] &=
  \asumnew[{\{\afishact[][Request][Requested]\}}][{\inp[][Requested]\inp[][Bid]\inp[][BidderID]{\afish[1][P]}}]
  \\
  \afish[1][P] &= \asumnew[{\{\afishact[][Select][Selected \logcat PassengerID]\}}]
  [{\inp[][Bid]\inp[][BidderID]\afish[1][P],\inp[][Selected]\inp[][PassengerID]\afish[2][P]}]
  \\
  \afish[2][P] &= \asumnew[{\{\afishact[][Cancel][Cancelled]\}}]
  [{\inp[][Cancelled]\inp[][Receipt]\zero,\inp[][Arrived]\afish[3][P]}]
  \\
\afish[3][P] &= \asumnew[{\{\afishact[][Start][Started]\}}][{\inp[][Started]\afish[4][P]}]
  \\
  \afish[4][P] &= \asumnew[{\{\afishact[][Finish][Finished]\}}]
  [{\areaction[][Path][{\afish[4][P]}],\inp[][Finished]\areaction[][Receipt][\zero]}]
  \\[1em]
  \text{taxi:}\quad\afish[][T] &= \inp[][Requested]\afish[1][T]
  \\
  \afish[1][T] &= \asumnew[{\{\afishact[][Offer][Bid]\}}]
  [{\inp[][Bid]\inp[][BidderId]\afish[1][T],\inp[][Selected]\inp[][PassengerID]\afish[2][T]}]
  \\
  \afish[2][T] &= \asumnew[{\{\afishact[][Arrive][Arrived]\}}]
  [{\inp[][Arrived]\inp[][Started]\afish[3][T],\inp[][Cancelled]\inp[][Receipt]\zero}]
  \\
  \afish[3][T] &= \asumnew[{\{\afishact[][record][Path]\}}]
  [{\inp[][Path]\afish[3][T],\inp[][Finished]\inp[][Receipt]\zero}]
  \\[1em]
  \text{office:}\quad\afish[][O] &= \inp[][Requested]\inp[][Bid]\afish[1][O]
  \\
  \afish[1][O] &= \asumnew[][{\inp[][Bid]\afish[1][O],\inp[][Selected]\afish[2][O]}]
  \\
  \afish[2][O] &= \asumnew[][{\inp[][Arrived]\inp[][Started]\afish[3][O],\inp[][Cancelled]\afish[4][O]}]
  \\
  \afish[3][O] &= \asumnew[][{\inp[][Path]\afish[3][O],\inp[][Finished]\afish[4][O]}]
  \\
  \afish[4][O] &= \asumnew[{\{\afishact[][Receipt][Receipt]\}}][{\inp[][Receipt]\zero}]
\end{align*}

The graphical representation of the machines as automata are shown in
\cref{ex:proj-well-formed}.
The passenger (like the taxis)
processes
all event types,
while the office skips events that are not required for following the overall progress of the protocol---in particular the office does not get to see the $\aeventtype[][PassengerID]$ event
that the taxi uses to find, identify, and pick up the passenger.

In the following we name events exactly like their types, adding indexes where further distinction is required.
For example, $\eventtyping[][Receipt_1][Receipt]$.

Starting with a swarm $\mksys{P[],T[],T[],O[]}\emptylog$---featuring two taxis that we name A and B---the only possible reduction is to invoke the $\acmd[][Request]$ command,
yielding the swarm $\mksys{P[\mklog{\aevent[][Requested]}],T[],T[],O[]}\mklog{\aevent[][Requested]}$.
Now the passenger is in state $\afish[1][P]$ and could already select a taxi,
but our passenger decides to wait for bids.
After propagation of the $\aevent[][Requested]$ event to both taxis
they may concurrently invoke their $\acmd[][Offer]$ command,
leading for example to the swarm
\begin{equation}\label{eq:2bids}
 \begin{array}{l}
   \mksys{P[\mklog{\aevent[][Requested]}],T[\mklog{\aevent[][Requested], \aevent[A][Bid], \aevent[A][BidderID]}], T[\mklog{\aevent[][Requested], \aevent[B][Bid], \aevent[B][BidderID]}],O[]
	\\\hfill}\mklog{\aevent[][Requested], \aevent[B][Bid], \aevent[B][Bidder], \aevent[A][Bid], \aevent[A][Bidder]}
 \end{array}
\end{equation}
Note that the propagation in~\eqref{eq:2bids} is not complete: events $\aevent[A][Bid]$ and $\aevent[A][BidderID]$ (resp. $\aevent[B][Bid]$ and  $\aevent[B][BidderID]$)) is only in the global log and the local logs of taxi A (reps. B); it has not been propagated to taxi B (resp. A).
Also, observe that the order of these events in the global log is \quo{arbitrary}; in fact the opposite order is also possible.

At this point a third taxi C enters the picture.
We model such dynamic behaviour by adding a new machine to the swarm that conceptually has been there from the beginning
but that has so far not been the target of a \rulename{Local} or \rulename{Prop} rule; this has no effect on the reductions performed on the smaller swarm because no rule requires us to inspect or modify all swarm participants.

As the request and bids propagate through the swarm, the following state can be reached:\begin{multline}
  \mksys{P[\mklog{\aevent[][Requested], \aevent[B][Bid], \aevent[B][Bidder]}],T[\mklog{\aevent[][Requested], \aevent[B][Bid], \aevent[B][Bidder], \aevent[A][Bid], \aevent[A][Bidder]}]} \\
  \mksys{T[\mklog{\aevent[][Requested], \aevent[B][Bid], \aevent[B][Bidder], \aevent[A][Bid], \aevent[A][Bidder]}],O[],T[\mklog{\aevent[][Requested], \aevent[A][Bid]}]} \\
  \mklog{\aevent[][Requested],\aevent[B][Bid],\aevent[B][Bidder], \aevent[A][Bid], \aevent[A][Bidder]}
\end{multline}
If now the passenger selects taxi B while concurrently taxi C places its bid we might end up with the global log
$\alog[\text{auc}]=\mklog{\aevent[][Requested], \aevent[B][Bid], \aevent[B][Bidder], \aevent[A][Bid], \aevent[A][Bidder], \aevent[][Selected], \aevent[C][Bid], \aevent[C][BidderID], \aevent[][passengerID]}$ and the swarm
\begin{multline}
  \mksys{P[\mklog{\aevent[][Requested], \aevent[B][Bid], \aevent[B][Bidder], \aevent[][Selected], \aevent[][PassengerID]}]} \\
  \mksys{T[\mklog{\aevent[][Requested], \aevent[B][Bid], \aevent[B][Bidder], \aevent[A][Bid], \aevent[A][Bidder]}], T[\mklog{\aevent[][Requested], \aevent[B][Bid], \aevent[B][Bidder], \aevent[A][Bid], \aevent[A][Bidder]}]} \\
  \mksys{O[],T[\mklog{\aevent[][Requested], \aevent[A][Bid], \aevent[A][Bidder], \aevent[C][Bid], \aevent[C][Bidder]}]}\alog[\text{auc}]
\end{multline}
At this moment the passenger thinks that there is a single bid followed by selection, the taxis assume two bids
(but different ones) without selection, and the office has no clue yet.
Once all events have been propagated, the passenger realises that there was a second bid,
everybody knows that the auction has ended, and taxi C finds that their bid has been ignored because it came too late.

With all machines respectively in states $\afish[2][P]$, $\afish[2][T]$, and $\afish[1][O]$
we find ourselves at a crossroads: either the ride happens or it is cancelled.
Since both the passenger and the taxi may act concurrently, the swarm may enter an inconsistent state.
Assuming concurrent invocation of $\acmd[][arrive]$ and $\acmd[][cancel]$
followed by propagation of the $\aevent[][cancelled]$ event to the office, the swarm is in state
\begin{multline}
  \mksys{P[\mklog{\alog[\text{auc}],\aevent[][Cancelled]}],T[\mklog{\alog[\text{auc}]}],T[\mklog{\alog[\text{auc}],\aevent[][Arrived]}]} \\
  \mksys{O[\mklog{\alog[\text{auc}],\aevent[][Cancelled]}],T[\mklog{\alog[\text{auc}]}]}\mklog{\alog[\text{auc}], \aevent[][Arrived], \aevent[][Cancelled]}
\end{multline}
Taxi A and C are still waiting for the choice to be made,
taxi B is waiting for the passenger to invoke the $\acmd[][Start]$ command,
the passenger is waiting for the office to invoke the $\acmd[][Receipt]$ command,
and the office will likely do just that.
After the office generates the $\aevent[][Receipt_1]$ event the global log necessarily is
$\alog[\text{confused}]=\mklog{\alog[\text{auc}], \aevent[][Arrived], \aevent[][Cancelled], \aevent[1][Receipt]}$.
Once the office disseminates all its local log to every other machine, all machines except for taxi B
will consider the protocol finished without a ride, with only taxi B being aware of the conflict:
\begin{multline}
  \mksys{P[\mklog{\alog[\text{auc}],\aevent[][Cancelled],\aevent[1][Receipt]}]} \\
  \mksys{T[\mklog{\alog[\text{auc}],\aevent[][Arrived],\aevent[][Cancelled],\aevent[1][Receipt]}],T[\mklog{\alog[\text{auc}], \aevent[][Cancelled], \aevent[1][Receipt]}]}\\
  \mksys{O[\mklog{\alog[\text{auc}],\aevent[][Cancelled],\aevent[1][Receipt]}],T[\mklog{\alog[\text{auc}],\aevent[][Cancelled],\aevent[1][Receipt]}]}\alog[\text{confused}]
\end{multline}

But in fact, as soon as taxi B manages to transmit the $\aevent[][Arrived]$ event for example to the passenger,
that machine recomputes its state based on the new local log $\mklog{\alog[\text{auc}], \aevent[][Arrived], \aevent[][Cancelled], \aevent[1][Receipt]}$
and ends up in state $\afish[3][P]$.
As a consequence, the ride can now begin with the $\acmd[][start]$ command.
If the passenger no longer wishes to travel with the taxi, they still need to follow the protocol of
starting and finishing a possibly empty ride, so that the office can generate a receipt for it
(there might be a fee for late cancellation).

\subsection{From regular trees to finite LTS}\label{sec:regtrees2lts}
Regular trees can be finitely represented as finite labelled transition systems.
We show how to translate terms in our syntax of machines (cf. \eqref{eq:lt}) in the graphical notation used in the paper.

A machine can equivalently be represented in the form of a finite labelled transition system, with $\acmdrel$ becoming self-loops and $\areaction$ becoming the transition from $\afish[]$ to $\afish[i]$.
More precisely, we can associate a finite-state automaton to a machine $\afish = \asumnew[@][{\areaction[1], \cdots, \areaction[n]}]$ as follows: the states of the automaton are the subterms of $\afish$; in state $\afish$ there is a self-loop transition to $\afish$ labelled $\acmd \cte \alogtype$ for each $\acmd \cte \alogtype \in \acmdrel$ and there is a transition labelled $\areaction[i]$ to state
$\afish[i]$ for each $i \in \natseg$.
And likewise for each sub-machine $\afish[i]$.
Intuitively, a branch $\areaction[i]$ describes a transition in which the machine consumes an event of type $\aeventtype[i]$ (dubbed \emph{guard}) from the log and continues in a state described by $\afish[i]$.
Note that this construction yields a finite-state automaton by the regularity of $\afish$.

Likewise, a finite-state automaton can be built out of a term of the syntax for swarm protocols give in \cref{sec:gt} as follows.
The state $\agt = \agsum$ has a transition to state $\agt[i]$ labelled with $\gsumprefix[i]$ for each $i \in \natseg$ representing that a participant of role $\arole[]$ executes a command $\acmd$ which produces a non-empty sequence of events of type $\alogtype$.
Note that this construction yields automata with multiple transitions out of a given state labelled with different roles; this corresponds to having non-local decisions as discussed in \cref{sec:gt}.

\subsection{Ensuring properties}\label{sec:problems}

The following examples are related to the problems in \cref{sec:causpropprob}.
\begin{example}[Causality]\label{ex:causal-dep}
  Consider a variation of the swarm protocol from the beginning of our running example,
  $\agt=\gsumprefix[][Request][Requested][P]\gtpref\gsumprefix[][Offer][Bid][T]\gtpref\agt'$.
The semantics of this type is that first \(\acmd[][Request]\) needs to happen,
  after which \(\acmd[][offer]\) occurs before the protocol continues.
If \(\reactsto(\arole[][P])=\emptyset\) then command \(\acmd[][Request]\) stays enabled forever, so in order to be faithful we must require \(\aeventtype[][requested]\in\reactsto(\arole[][P])\).
Similarly, we need to require \(\aeventtype[][Bid]\in\reactsto(\arole[][T])\)
  lest the taxi machine never reach the final state and keep emitting events.
But $\arole[][T]$ must also react to \(\aeventtype[][Requested]\),
  otherwise \(\acmd[][Offer]\) may be invoked concurrently with \(\acmd[][Request]\)
  and the merged log could be $\mklogtype{\aevent[][Bid], \aevent[][Requested]}$, violating the expected semantics.
  \finex
\end{example}

\begin{example}[Propagation]\label{ex:bad-propagation}
  Consider the swarm protocol
  \begin{align*}
    \agt&=
    \gsumprefix[][Select][Selected,PassengerID][P]\gtpref
    \gsumprefix[][Arrive][Arrived][T]\gtpref
    \gsumprefix[][Receipt][Receipt][O]
  \end{align*}
  and the subscription
  \begin{align*}
    \reactsto&:
    \begin{cases}
      \arole[][P]\mapsto\{\aeventtype[][Selected],\aeventtype[][Arrived]\} \\
      \arole[][T]\mapsto\{\aeventtype[][Selected],\aeventtype[][Arrived]\} \\
\arole[][O]\mapsto\{\aeventtype[][Selected],\aeventtype[][PassengerID],\aeventtype[][Arrived],\aeventtype[][Receipt]\}
	 \end{cases}
  \end{align*}
  We obtain the following projections:
  \begin{align*}
    \afish[{\arole[][P]}] = \agt\aproj[{\arole[][P]}] &=
    \asumnew[{\{\afishact[][Select][Selected\logcat PassengerID]\}}][{\inp[][Selected]\inp[][Arrived]\zero}]
    \\
    \afish[{\arole[][T]}] = \agt\aproj[{\arole[][T]}] &=
    \inp[][Selected]\asumnew[{\{\afishact[][Arrive][Arrived]\}}][{\inp[][Arrived]\zero}]
    \\
    \afish[{\arole[][O]}] = \agt\aproj[{\arole[][O]}] &=
    \inp[][Selected]\inp[][PassengerID]\inp[][Arrived]\asumnew[{\{\afishact[][Receipt][Receipt]\}}][{\inp[][Receipt]\zero}]
  \end{align*}
  Starting with an empty log the only possible reduction is to invoke \(\acmd[][Select]\),
  producing the log $\mklog{\aevent[][Selected], \aevent[][PassengerID]}$.
  If we now let only \(\aevent[][Selected]\) propagate to \(\afish[{\arole[][T]}]\) we may invoke \(\acmd[][Arrive]\) to produce event $\aevent[][Arrived]$.
  Since event \(\aevent[][PassengerID]\) is not known at \(\afish[{\arole[][T]}]\)
  the log merge operation may order \(\aevent[][Arrived]\) before \(\aevent[][PassengerID]\).
  At this point \(\afish[{\arole[][O]}]\) is blocked: its command \(\acmd[][Receipt]\) will never be enabled because
  event \(\aevent[][Arrived]\) has been ignored
  (recall that \(\stchange(\afish[{\arole[][O]}],\mklog{\aevent[][Selected], \aevent[][Arrived], \aevent[][PassengerID]})=
  \inp[][Arrived]\asumnew[{\{\afishact[][Receipt][Receipt]\}}][{\inp[][Receipt]\zero}]\)).
\finex
\end{example}

\begin{example}\label{ex:causal-dep-correct}
  Note how the rules for causality require precisely the subscriptions discussed in \cref{ex:causal-dep}.
Also  the rules require role $\arole[][T]$ to subscribe  to the event type $\aeventtype[][PassengerID]$
  to fix the issue with \cref{ex:bad-propagation}.
  \finex
\end{example}

The following examples is an instance of the problem in \cref{sec:choiceprobs}.
\begin{example}[Divergent local choice]\label{ex:bad-choice}
  Consider the main choice in our running example (with irrelevant path tracking removed):
  \begin{align*}
    \agt[\text{choose}] =\ &
    \gsumprefix[][Arrive][Arrived][T]\gtpref\gsumprefix[][Finish][Finished][P]\gtpref\gsumprefix[][Receipt][Receipt][O]\zero \\
    &+\gsumprefix[][Cancel][Cancelled][P]\gtpref\gsumprefix[][Receipt][Receipt][O]\zero
  \end{align*}
  Assume that \(\acmd[][Arrive]\) and \(\acmd[][Cancel]\) are invoked concurrently such that the resulting global log is $\mklog{\aevent[][Arrived], \aevent[][Cancelled]}$.
  The restrictions of \cref{def:cc} only require \(\reactsto(\arole[][O])\supseteq\{\aeventtype[][Finished],\aeventtype[][Cancelled],\aeventtype[][Receipt]\}\).
In this case with the aforementioned log a machine of role \(\arole[][O]\) incorrectly assumes that the lower branch was taken
  while the swarm protocol proceeds with the upper branch.
\finex
\end{example}

The issue in \cref{ex:bad-choice} is due to the fact that role \(\arole[][O]\) is oblivious of a branch of the choice even though it is still involved in the continuation of such branch.

The following examples is an instance of the problem in \cref{sec:interference}.
\begin{example}[Ambiguous decision]\label{ex:confused-choice}
  Suppose that \(\reactsto=\roleset\times\{\typeset\}\) with the following variation of our running example:
  \begin{align*}
    \agt[\text{choose}] =\ &
    \gsumprefix[][Arrive][Arrived][T]\gtpref\gsumprefix[][Finish][Finished,Rating][P]\gtpref\gsumprefix[][Receipt][Receipt][O]\zero \\
    &+\gsumprefix[][Cancel][Finished][P]\gtpref\gsumprefix[][Receipt][Receipt][O]\zero
  \end{align*}
  If $\acmd[][Arrive]$ and $\acmd[][Cancel]$ are invoked concurrently
  then the resulting log may be $\mklog{\aevent[][Arrived], \aevent[][Finished]}$
  which would inhibit further progress:
  command $\acmd[][Finish]$ can no longer be enabled so the $\aeventtype[][Rating]$ event cannot be emitted.
Even without this deadlock, we note that the upper branch would process a $\aeventtype[][Finished]$ event originating from the losing lower branch.
\finex
\end{example}

\subsection{Anomalies}\label{sec:anomalies}
For readability we repeat the anomalies described at the end of \cref{sec:pieces}
\begin{itemize}
\item Machines could have commands enabled that would be disabled if
  the model were synchronous; this may lead to the emission of events
  that need to be ignored later.
\item Events are ignored according to their type only, therefore even
  after full propagation of the events in the global log a machine may
  process events stemming from the anomalous invocation of a command.
\end{itemize}
The first point is aptly illustrated by the running example, in
particular in \cref{sec:full-example}.
We illustrate the second point on the swarm protocol $\agt=\gsumprefix[][c][t]\gtpref\gsumprefix[][c][t]\gtpref\zero$
with $\reactsto(\arole[])=\{\aeventtype[]\}$ and $\afish=\agt\aproj =
  \asumnew[{\{\afishact[][c][t]\}}][{\inp[][t]\asumnew[{\{\afishact[][c][t]\}}][{\inp[][t]\zero}]}]
$.
Consider the swarm $\mksys{M[],M[]}\emptylog$ and let both machines concurrently invoke command $\acmd$, leading to the swarm $\mksys{M[\mklog{\aevent[1][t]}],M[\mklog{\aevent[2][t]}]}\mklog{\aevent[1][t],\aevent[2][t]}$.
Propagating the global log to both machines will result in both of them reaching the final state, even though they should have ignored the second event since it came from an anomalous invocation of $\acmd$.

While the above could be avoided by forbidding multiple instances of the same role, a similar anomaly cannot be fixed in the same way.
Consider the universal subscription $\reactsto=\roleset\times\{\typeset\}$,
the swarm protocol $\agt=\gsumprefix[][c_1][a\logcat c][A]\gtpref\zero + \gsumprefix[][c_2][b\logcat c][B]\gtpref\zero$, the projections $\afish[][A]=\agt\aproj[{\arole[][A]}]$ and $\afish[][B]=\agt\aproj[{\arole[][B]}]$, and the swarm $\mksys{A[],B[]}\emptylog$.
Concurrent invocation of both commands may well yield the global log $\mklog{\aevent[][b], \aevent[][a], \aevent[1][c], \aevent[2][c]}$ with $\source(\aevent[][c_1])=1$.
Propagation to all machines will have them interpret $\aevent[][b]$, discard $\aevent[][a]$,
and then process $\aevent[][c_1]$ instead of $\aevent[][c_2]$---the latter would be more fitting because it is the companion to $\aevent[][b]$ originating from $\acmd[2]$.

\subsection{Proofs of \cref{sec:fishes,sec:wf,sec:correct}}

\restateWfPreservation*
\begin{proof} It follows by induction on the derivation
  $(\asysrt,\alog)\red[\alpha](\asysrt',\alog')$ .  \rulename{Prop} is immediate because
  of the premise $\alog[i]'\sublog\alog$ while rule \rulename{Local}
  holds by the definition of $\_\mergelog\_$.
\end{proof}

\restateEvCons*
\begin{proof} By  induction on $n$ after noting that  $(\asysrt, \alog) \red[\tau] (\upd \asysrt i {(\afish[i], \alog)}, \alog)$
holds for all $i$ by rule \rulename{Prop}.
\end{proof}

\restateProjdet*

\begin{proof} The proof follows by coinductive arguments and by noting that \cref{def:cd} ensures each role to be subscribed to the first type in the emitted events, which should be different by \cref{def:wb}.
\end{proof}

\restateCmdEnabledAdmissibleLog*
\begin{proof}
Assume $\agt = \agsum$ and proceed by induction on $\alog$.
Base case follows straightforwardly by
  taking $\alog' = \emptylog$. For the inductive step we consider
  $\alog = \aevent\logcat\alog'$ and proceed by cases.

  If $\efftype[{\aevent}] = \emptylog$ then $\eventtyping$ and
  $\aeventtype\not\in\rdy[{\agt\aproj}]$. Hence,
  $\stchange(\agt\aproj,\aevent\logcat\alog')
  =\stchange(\agt\aproj,\alog')$, and the case follows by inductive
  hypothesis.

  If $\efftype [{\aevent}]= \aeventtype$. By definition of effective
  type,
there exists $i \in I$ such that
  $\agt \red[{\gsumprefix[i]}] \agt'' = \agt[i]$ such
  that
  $\filter[{\alogtype[i],\reactsto[{\arole[]}]}] =
  \aeventtype\logcat\alogtype[i]'$.  Then, there are two cases.
  \begin{itemize}
  \item[(i)] If there exist $\alog[1]$ and $\alog[2]$ such that
    $\alog' = \alog[1]\logcat\alog[2]$ and
    $\efftype [{\alog[1]}]= \alogtype[i]'$. Then,
    $\obscmd[{\stchange(\agt\aproj,\alog)}]$ iff
    $\obscmd[{\stchange(\agt''\aproj,\alog[2])}]$. By inductive
    hypothesis, we conclude that there exists
    $\eqlog{\alog[2]'}{\alog[2]}$ such that
    $(\agt'',\emptylog) \weakred[] (\agt',\alog[2]')$ and
    $\Stchange(\agt'',\alog[2]) \red[{\gsumprefix[][@][{\alogtype}]}]
    \agt' $.  The proof is completed by taking
    $\alog' = \alog[1]'\logcat\alog[2]'$ with $\alog[1]'$ any fresh
    log of type $\alogtype[i]$.
  \item[(ii)] If for all $\alog[1]$ and $\alog[2]$ such that
    $\alog = \alog[1]\logcat\alog[2]$ we have that
    $\efftype [{\alog[1]}] = \alogtype[1]\neq \alogtype[i]$. Then, we
    conclude that $\efftype [\alog]$ is a proper prefix of
    $\alogtype[i]$; consequently,
    $\stchange(\agt\aproj,\alog) = \inp[1] \afish$, which contradicts
    the hypothesis that $\obscmd[\stchange(\agt\aproj,\alog)]$.
  \end{itemize}
\end{proof}

\restateAdmissibleReduction*

\begin{proof}
By induction on the length of $\efftype = \alogtype$. Base case follows straightforwardly. For the inductive step,
consider $ \alogtype =  \alogtype' \logcat \aeventtype$. By definition of $\efftype$, there should exist
$\aevent\in\alog$ such that $\eventtyping$ and
$\agt \red[{\gsumprefix[][@][{\alogtype[e]}]}] \agt'$ with $\aeventtype \in \alogtype[e]$.
Since $\alog$ is admissible, its effective type coincides with the effective type of a log obtained by the
merging of several executions of $\agt$, i.e., $\{(\agt,\emptylog) \weakred[] (\agt,\alog[i][m])\}_{i\in \natseg[k]}$ (hence, all relevant events of $\alogtype[e]$ are in the log).
Therefore, there exists $\alogtype[1]$ such that
$\alogtype = \alogtype[1]\logcat \filter[{\alogtype[e],\reactsto[{\activer[\agt']}]}]$
(the case $\agt' = \zero$ is analogous).
Consequently, there exist $\alog[1]$ and $\alog[2]$ such that  $\alog = \alog[1]\logcat\alog[2]$, and
$\alogtype[1] = \efftype[{\alog[1]}][\agt]$ and
$\efftype[{\alog[2]}][\agt'] = \filter[{\alogtype[e],\reactsto[{\activer[\agt']}]}]$.
It should be noted that $\alog[1]$ is admissible for $\agt$, since it can be matched with a log from
$\alog' \in (\mergelog_{i\in \natseg[k]} \alog[i]')$ where
$(\agt,\emptylog) \weakred[] (\agt,\alog[i]')\weakred[] (\agt,\alog[i])$ and
$\alog[i]'\cap\alog[1] \neq \emptyset$ for all $i \in \natseg[k]$.
By inductive hypothesis,
there exists $(\agt,\emptylog) \weakred[] (\agt,\alog[1]')$ with $\eqlog{\alog[1]'}{\alog[1]}$. Then,
the proof is completed by taking $\alog' = \alog[1]' \logcat \alog[2]'$ where $\alog[2]'$ is any fresh log of the appropriate type, i.e.,
$\eventtyping[][{\alog[2]'}][{\alogtype[e]}]$.
\end{proof}

\restateAdmissibleMerge*
\begin{proof}
  Since $\alog[1]$ is admissible for $\agt$, there exist
  $\{(\agt,\emptylog) \weakred[] (\agt,\alog[i][m])\}_{i\in
    \natseg[k]}$, and
  $\alog[1]' \in (\mergelog_{i\in \natseg[k]} \alog[i][m])$ with
  $\eqlog{\alog[1]'}{\alog[1]}$.
Since $\alog[2]$ is admissible,
$(\agt,\emptylog) \weakred[] (\agt,\alog[2]')$ by \cref{lem:admissible-reduction}. Hence, take $\alog[k+1][m] = \alog[2]'\logcat\alog[3]$.
Then,
 $(\agt,\emptylog) \weakred[] (\agt,\alog[k+1][m])$.
 Take
 $\alog \in (\mergelog_{i\in \natseg[k+1]} \alog[i][m])$.
Assume $\alog[3] = \aevent[1]\ldots\aevent[n]$. Then,
$\alog = \alog[0][h]\logcat\aevent[1]\logcat\alog[h]'$ where
$\alog[h]'\in \aevent[2]\ldots\aevent[n] \mergelog \alog[1][h]$ with
$\alog[1] = \alog[0][h]\logcat \alog[1][h]$
and $\alog[2]\subseteq\alog[0][h]$.
Then $\efftype=  \alogtype[0][h] \logcat \efftype[{\aevent[1]\logcat\alog[h]'}][{\agt[0]}]$.
There are two cases.
 If $\agt[0]\neq\Stchange(\agt,\alog[2])$, i.e., different from the global graph that enables the execution of
 $\acmd$, then $\efftype[{\aevent[1]\logcat\alog[h]'}][{\agt[0]}] = \efftype[{\alog[h]'}][{\agt[0]}]$, i.e., the event is not reflected in the type. Moreover,  the remaining events in $\aevent[2]\ldots\aevent[3]$ cannot be taken as guards because of invariance with respect to guards; hence they cannot change the effective type. Consequently,  $\efftype = \efftype[{\alog[1]}]$.

 If $\agt[0] =\Stchange(\agt,\alog[2])$, then $\efftype=  \alogtype[0][h] \logcat \efftype[{\aevent[1]\logcat\alog[h]'}][{\agt[0]}] =
 \alogtype[0][h] \logcat\  \filter[{\alogtype[e],\reactsto[{\activer[\agt']}]}] \logcat \efftype[{\alog[h]''}][{\agt'}]$ with
 $\alog[h]''\in \aevent[k]\ldots\aevent[n] \mergelog \alog[1][h]''$ and $k>1$ and  $\alog[1][h]= \alog[1][h]' \logcat \alog[1][h]''$. Then, there are two cases, either $\efftype[{\alog[h]''}][{\agt'}] = \emptylog$  or $\efftype[{\alog[h]''}][{\agt'}] = \aeventtype[e] \logcat \alogtype[h]'''$. The former case follows immediately; the later follows by notting that an event of type $\aeventtype[e]$ should be in $\alog[1][h]''$ because none of $\aevent[k]\ldots\aevent[n]$ for $k>0$ can have type $\aeventtype[e]$ ($\aeventtype[e]$ is a guard of $\agt'$ and by invariance it cannot appear in two transitions).
\end{proof}

\restateSerializability*

\begin{proof}
  The proof follows by induction on the length of the derivation.
The base case follows immediately.
For the inductive case, take
  $(\asysrt,\emptylog) \weakred (\asysrt'',\alog'') \red[\alpha]
  (\asysrt',\alog)$.
By inductive hypothesis, $\alog''$ is admissible for
  $\agt$.
Then, we proceed by case analysis on the rule for the last
  reduction:
  \begin{itemize}
  \item {\sc Local}. Then, there exists $i \in \dom \asysrt$ such that
    $\asysrt(i) = (\agt\aproj[{\arole}],\alog[i])$, and
    $\agt\aproj[{\arole[]}]\ \alog[i]
    \red[{\afishact[][@][\alogtype]}] \agt\aproj[{\arole[]}]\
    \alog[i]'$, and $\alog \in \alog'' \mergelog \alog[i]'$. By
    \cref{lem:cmd-on-admissible-log}, $\alog[i]$ is admissible.
By rule {\sc Cmd}, $\alog[i]' = \alog[i]\logcat\alog[i]''$ for a
    fresh $\alog[i]''$ such that
    $\eventtyping[][{\alog[i]'}][\alogtype]$. It is immediate that
    $\alog[i]'$ is admissible. Then,
    $\alog \in \alog'' \mergelog \alog[i]'$ is admissible by
    \cref{merge-admissible}.
  \item {\sc Prop}. Then, $\alog = \alog''$ which is admissible for
    $\agt$.
  \end{itemize}
\end{proof}

\restateProjMimick*

\begin{proof} By induction on the length $k$ of the run.
Case $k=0$ is immediate. For the inductive step, we assume
$(\agt,\emptylog) \weakred[] (\agt,\alog') \red[\afishact] (\agt,\alog)$.
By rule {\sc G-Cmd},  $\Stchange(\agt, \alog')=\agt' \red[\afishact] \agt''$ and
$\alog = \alog' \logcat \alog''$ with  $\eventtyping[][\alog''][\alogtype]$.
From $\agt' \red[\afishact] \agt''$, we conclude that
$\agt' = \agsum$ and there exists $j\in I$ such that
$\acmd[j] = \acmd$ and $\alogtype[j] = \alogtype$ and $\agt[j] = \agt''$. Then,
\begin{align*}
\Stchange(\agt, \alog) = \ & \Stchange(\agt, \alog'\logcat\alog'') & \text{because }\ \alog = \alog' \logcat \alog' \\
=\ &  \Stchange(\agt', \alog'') & \text{because }\ \agt'=\Stchange(\agt, \alog')\\
=\ & \agt'' = \agt[j] & \text{because}\ \agt'=\agsum \ \text{and}\ \eventtyping[][\alog''][\alogtype] \\
\end{align*}
We need to show that  $\stchange(\agt\aproj,\alog' \logcat \alog'') = \agt''\aproj$.
By inductive hypothesis, $\stchange(\agt\aproj,\alog') = \stchange(\agt,\alog')\aproj = \agt'\aproj$.
Then,   $\stchange(\agt\aproj,\alog' \logcat \alog'') = \stchange(\agt'\aproj,\alog'')$.
Note that
\[\agt'\aproj =  (\agsum)\aproj  = \asumnew[@][{(\gsumprefix \gtpref \agt[i]) \aproj \sst i \in I }]\]

We have two cases, if $\arole[]\not\in\roles[\agt']$, then $\agt'\aproj = \zero$ and $\agt[i]\aproj = \zero$ for all
$i\in I$. Hence,

 $$\stchange(\agt\aproj,\alog' \logcat \alog'') = \stchange(\agt'\aproj, \alog'')  = \zero =
\agt[j]\aproj =\  \agt''\aproj = \Stchange(\agt, \alog)\aproj$$

 In case, $\arole[]\in\roles[\agt']$, by \cref{def:wb} then $\hd {{\alogtype[j]}} \in \reactsto(\arole[])$ and $(\gsumprefix[j])\aproj \neq \zero$. Since $\eventtyping[][\alog''][{\alogtype[j]}]$,
 it is immediate that
 there exists $\alog[1]$ and $\alog[2]$ s.t. $\alog'' = \alog[1]\logcat\alog[2]$ and
$$\stchange({\filter[{\alogtype[j], \reactsto[{\arole[]}]}]\ {\colorSymb ?}.\ (\agt[j]\aproj)  , \alog''}) =
 \stchange({\agt[j]\aproj, \alog[2]})$$
that is, all events matching $\filter[{\alogtype[j], \reactsto[{\arole[]}]}]$ are processed. It should be noted
that none of the events in $\alog[2]$ have type in $\reactsto[{\arole[]}]$; consequently,
$ \stchange({\agt[j]\aproj, \alog[2]}) = \agt[j]\aproj = \agt''\aproj$.
\end{proof}
\restateRealisationCovers*
\begin{proof}
By induction on the length of the reduction. Base case follows straightforwardly. The inductive step follows by inductive hypothesis, \cref{lem:projpres} and by propagating all events to all replicas.
\end{proof}

\end{document}